\documentclass[a4paper,twoside,final,11pt,onecolumn]{PAC51elsart}

\usepackage{color}
\usepackage{graphicx}
\usepackage{latexsym}
\usepackage{amssymb,amsmath,amstext,bm}
\usepackage{multicol,multirow}
\usepackage{rotating}
\usepackage{longtable}
\usepackage{colordvi}
\usepackage{ulem}
\usepackage[numbers]{natbib}
\usepackage{epsfig}
\usepackage{graphicx}
\usepackage{subfigure}
\usepackage{marvosym}
\usepackage{subfigure}
\usepackage[dvipsnames]{xcolor}
\usepackage{siunitx}
\usepackage{pstricks}
\usepackage{fancyhdr}
\usepackage{subfigure}
\usepackage{lscape}
\usepackage{afterpage}
\usepackage{enumitem}
\usepackage{hyperref}

\newsavebox{\Imagebox}

\newcommand{\emep}{$e^{-}e^{+}\;$}

\begin{document}

%

\hyphenation{brems-strah-lung}
\hyphenation{po-si-trons}

%
\pagestyle{empty}

\begin{frontmatter}

{\huge Proposal to PAC51}

\vspace*{7.5pt}

{\huge PR12+23-002}

\vspace*{7.5pt}

\title{\bf\LARGE{Beam Charge Asymmetries for} \\ \vspace*{8pt} \bf\LARGE{Deeply Virtual Compton Scattering} \\ \vspace*{8pt} \bf\LARGE{on the Proton at CLAS12}}

\vspace*{2.5pt}
\author {\sf V.~Burkert$^{\star}$, }
\author {\sf A.~Camsonne, }
\author {\sf P.~Chatagnon, }
\author {\sf A.~Deur, }
\author {\sf L.~Elouadrhiri, }
\author {\sf F.-X.~Girod, }
\author {\sf J.~Grames, }
\author {\sf D.~Higinbotham, }
\author {\sf V.~Kubarovsky, }
\author {\sf M.~McCaughan, }
\author {\sf R.~Paremuzyan$^{\star}$, }
\author {\sf E.~Pasyuk, }
\author {\sf M.~Poelker, }
\author {\sf X.~Wei }
\vspace*{-4pt}
\address{Thomas Jefferson National Accelerator Facility \\
12000 Jefferson Avenue, Newport News, VA 23606, USA}

\vspace*{4pt}

\author {\sf J.-S.~Alvarado-Galeano, }
\author {\sf M.~Atoui, }
\author {\sf R.~Dupr\'e, }
\author {\sf S.~Habet, }
\author {\sf A.~Hobart, }
\author {\sf D.~Marchand, }
\author {\sf D.~Matamoros, }
\author {\sf C.~Mu\~noz Camacho, }
\author {\sf S.~Niccolai$^{\star}$, }
\author {\sf M.~Ouillon, }
\author {\sf N.~Pilleux, }
\author {\sf E.~Voutier$^{\star\dag}$, }
\author {\sf P.-K.~Wang }
\vspace*{-4pt}
\address{Laboratoire de Physique des 2 Infinis Ir\`ene Joliot-Curie \\
Universit\'e Paris-Saclay, CNRS/IN2P3, IJCLab \\ 
15 rue Georges Cl\'emenceau, 91405 Orsay cedex, France}

\vspace*{4pt}

\author {\sf R.~Capobianco$^1$, } 
\author {\sf S.~Diehl$^{1,2}$, }
\author {\sf K.~Joo$^1$, }
\author {\sf A.~Kim$^1$, }
\author {\sf V.~Klimenko$^1$, }
\author {\sf R.~Santos$^1$, }
\author {\sf P.~Stoler$^1$} 
\vspace*{-4pt}
\address{$^1$University of Connecticut, Department of Physics \\
2152 Hillside Road, Storrs, CT U-3046, USA}

\vspace*{1pt}

\address{$^2$Universit\"at Gie\ss en \\
Luwigstra\ss e 23, 35390 Gie\ss en, Deutschland}

\vspace*{4pt}

\author {\sf B.~Raue}
\vspace*{-4pt}
\address{Florida International University \\ 
Modesto A. Maidique Campus \\
11200 SW 8th Street, CP 204, Miami, FL 33199, USA}

\vspace*{4pt}

\author {\sf M.~Battaglieri, }
\author {\sf A.~Celentano, }
\author {\sf R.~De Vita, }
\author {\sf L.~Marsicano, }
\author {\sf M.~Osipenko, }
\author {\sf M.~Ripani, }
\author {\sf M.~Spreafico, }
\author {\sf M.~Taiuti }
\vspace*{-4pt}
\address{Istituto Nazionale di Fisica Nucleare \\ 
Sezione di Genova e Dipatimento di Fisica dell'Universit\`a \\
Via Dodecaneso, 33 - 16146 Genova, Italia}

\vspace*{30pt}

{\small
\leftline{\it $^{\star}$Spokesperson}
\par
\leftline{\it $^{\dag}$Contact person}
}

\newpage

\author {\sf M.~Bondi}
\vspace*{-4pt}
\address{Istituto Nazionale di Fisica Nucleare \\ 
Sezione di Catania \\
Via Santa Sofia, 64 - 95123 Catania, Italia}

\vspace*{4.0pt}

\author {\sf A.~Bianconi$^{3,4}$, }
\author {\sf G.~Costantini$^{3,4}$, }
\author {\sf G.~Gosta$^{4}$, }
\author {\sf M.~Leali$^{3,4}$, }
\author {\sf V.~Mascagna$^{3,4}$, }
\author {\sf S.~Migliorati$^{3,4}$, }
\author {\sf L.~Venturelli$^{3,4}$}

\address{$^3$Universit\`a degli Studi di Brescia \\
Via Branze, 38 - 25121 Brescia, Italia}

\vspace*{1pt}

\address{$^4$Istituto Nazionale di Fisica Nucleare \\ 
Sezione di Pavia \\ 
Via Agostino Bassi, 6 - 27100 Pavia, Italia}

\vspace*{4pt}

\author {\sf L.~Barion$^3$, } 
\author {\sf G.~Ciullo$^{3,4}$, }
\author {\sf M.~Contalbrigo$^3$, }
\author {\sf P.~Lenisa$^{3,4}$, }
\author {\sf A.~Movsisyan$^3$, }
\author {\sf L.~Pappalardo$^{3,4}$ }
\vspace*{-4pt}
\address{$^3$Istituto Nazionale di Fisica Nucleare \\ 
Sezione di Ferrara \\ 
Via Saragat, 1 - 44122 Ferrara, Italia}

\vspace*{1pt}

\address{$^4$Universit\`a di Ferrara \\
Via Ludovico Ariosto, 35 - 44121 Ferrara, Italia}

\vspace*{4pt}

\author {\sf B.~Pasquini } 
\vspace*{-4pt}
\address{Universit\`a degli Studi di Pavia \\
Corso Strada Nuova, 65 - 27100 Pavia, Italia}

\vspace*{4pt}

\author {\sf P.L.~Cole}
\vspace*{-4pt}
\address{Lamar University, Physics Department \\
4400 MLK Boulevard, Beaumont, TX 77710, USA}

\vspace*{4pt}

\author {\sf Z.~Zhao } 
\vspace*{-4pt}
\address{Duke University \\ 
120 Science Drive, Durham, NC 27708, USA}

\vspace*{4pt}

\author {\sf P.~Gueye} 
\vspace*{-4pt}
\address{Michigan State University \\
640 South Shaw Lane, East Lansing, MI 48824, USA}

\vspace*{4pt}

\author {\sf M.~Defurne, } 
\author {\sf D.~Sokhan$^5$} 
\vspace*{-4pt}
\address{Institut de Recherche sur les Lois Fondamentales de l'Univers \\
Commissariat \`a l'Energie Atomique, Universit\'e Paris-Saclay \\
91191 Gif-sur-Yvette, France}

\vspace*{4pt}

\author {\sf B.~McKinnon}
\vspace*{-4pt}
\address{$^5$University of Glasgow \\
University Avenue, Glasgow G12 8QQ, United Kingdom}

\newpage

\author {\sf I.~Fernando} 
\vspace*{-4pt}
\address{University of Virginia \\ 
382 McCormick Road, Charlottesville, VA 22904, USA}

\author {\sf T.~Chetry } 
\vspace*{-4pt}
\address{Mississippi State University \\ 
355 Lee Boulevard, Mississippi State, MS 39762, USA}

\author {\sf M.~Hattawy, }
\author {\sf C.E.~Hyde}
\vspace*{-4pt}
\address{Old Dominion University \\
5115 Hampton Boulevard, Norfolk, VA 23529, USA}

\vspace*{4pt}

\author {\sf M.H.~Mitra}
\vspace*{-4pt}
\address{University of West Florida \\
11000 University Pkwy, Pensacola, FL 32514, USA}

\vspace*{4pt}

\author {\sf T.~Forest}
\vspace*{-4pt}
\address{Idaho State University \\
921 South 8th Avenue, Pocatello, ID 83209, USA}

\vspace*{4pt}

\author {\sf J.C.~Bernauer, } 
\author {\sf R.~Singh} 
\vspace*{-4pt}
\address{Stony Brook University \\
100 Nicolls Road, Stony Brook, NY 11794, USA}

\vspace*{4pt}

\author {\sf A.~Filippi}
\vspace*{-4pt}
\address{Istituto Nazionale di Fisica Nucleare \\ 
Sezione di Torino \\ 
Via P. Giuria, 1 - 10125 Torino, Italia}

\vspace*{4pt}

\author {\sf A.~Afanasev, W.J.~Briscoe, A.~Schmidt, I.~Strakovsky}
\vspace*{-4pt}
\address{The George Washington University \\
221 I Street NW, Washington, DC 20052, USA}

\vspace*{40pt}

{\bf\Large{a CLAS Collaboration and}}

\vspace*{20pt}

{\bf\Large{Jefferson Lab Positron Working Group}}

\vspace*{20pt}

{\bf\Large{Proposal}}

\vspace*{70pt}

{\small 22 May 2023}

%
%
\newpage

\null\vfill

\begin{abstract}
{\small
The parameterization of the nucleon structure through Generalized Parton Distributions (GPDs) shed a new light on the  nucleon internal dynamics. For instance, GPDs provide an unprecedented experimental access to the orbital momentum of the nucleon and the distribution of forces experienced by partons. For its direct interpretation, Deeply Virtual Compton Scattering (DVCS) is the golden channel for GPDs investigation. The DVCS process interferes with the Bethe-Heitler (BH) mechanism to constitute the leading order amplitude of the $eN \to eN\gamma$ process. The study of the $ep\gamma$ reaction with polarized positron and electron beams gives a complete set of unique observables to unravel the different contributions to the $ep \gamma$ cross section. This separates the different reaction amplitudes, providing a direct access to their real and imaginary parts, which greatly simplifies physics interpretation. This procures crucial constraints on the model dependences and associated systematic uncertainties on GPDs extraction. The real part of the BH-DVCS interference amplitude is particularly sensitive to the $D$-term which parameterizes the Gravitational Form Factors of the nucleon. The separation of the imaginary parts of the interference and DVCS amplitudes provides insights on possible higher-twist effects.

We propose to measure the unpolarized and polarized Beam Charge Asymmetries (BCAs) of the $\vec{e}^{\pm}p \to e^{\pm}p \gamma$ process on an unpolarized hydrogen target with {\tt CLAS12}, using polarized positron and electron beams at 10.6~GeV. The azimuthal and $t$-dependences of the unpolarized and polarized BCAs will be measured over a large $(x_B,Q^2)$ phase space using a 2400-hour run with a luminosity of 0.66$\times 10^{35}$~cm$^{-2}\cdot$s$^{-1}$.
}
\end{abstract}

\end{frontmatter}
\vfill\eject
%
%
\pagestyle{plain}

\pagenumbering{roman}

\setcounter{page}{1}

%
%
%
%
%

\begin{center}
\LARGE{\bf Executive Summary \\}
\large{$\mathbf{\sim . \sim . \sim}$}
\end{center}

The Jefferson Laboratory has been a world center for the exploration of the internal structure and dynamics of nucleons and nuclei for over 25 years: first with the 6~GeV energy reach of the original Continuous Electron Beam Accelerator Facility (CEBAF), and for the past several years with the energy-upgraded 12~GeV electron accelerator and the new and upgraded equipment in the experimental end stations. Nearly all experiments have been carried out with the extremely precise electron beam of CEBAF. Some of the high impact science topics can be most cleanly explored in comparison of measurements carried out with electrons and with positrons. The recent development and successful test of a spin polarized positron source at Jefferson Lab has opened up a new line of measurements that complement the electron-induced measurements and lead to new insights into the structure of matter. 

The main physics goal of the present proposal is the measurement of the real part of the Compton Form Factor (CFF) $\mathcal{H}$ over the large kinematical domain accessible with the {\tt CLAS12} spectrometer. We propose to carry out measurements of Beam Charge Asymmetries (BCAs) in the Deeply Virtual Compton Scattering (DVCS) reaction on an  unpolarized hydrogen target using electron and positron beams produced at the future Ce$^+$BAF injector and accelerated through CEBAF up to 10.6~GeV. These secondary beams with an intensity of 50~nA and a polarization of 60\% will interact within a new hydrogen target cell featuring a larger diameter to accomodate for larger beam emittances. The existing M{\o}ller polarimeter upstream of the {\tt CLAS12} will be upgraded to allow operation for both electron and positron beams based on M{\o}ller and Bhabha scatterings.

The comparison of DVCS measurements with electrons $e^-(p,e^-p\gamma)$ and with positrons $e^+(p,e^+p\gamma)$ isolates the quantum interference amplitude between the Bethe-Heitler and DVCS processes. As such it provides the cleanest, model-independent access to the real part of a complex CFF, without the need for additional theoretical assumptions in the extraction procedure. The imaginary part can be directly accessed in the beam spin asymmetry employing the highly polarized electron or positron beam. 
Full knowledge of the real and imaginary parts of the CFF enables employing a dispersion relation that allows to determine a new form factor which is at the base of a new line of research to access the mechanical or gravitational properties of the nucleon. These particle properties can be directly measured only in the interaction of gravity with matter, which is experimentally a highly impractical proposition.

This BCA experiment was initially proposed to the Program Advisory Committee PAC46 as the Letter-of-Intent LOI12-18-004 followed by the full proprosal PR12-20-009 to PAC48. The proposal was Conditionally Approved C2 with some concerns about the impact of positron beams and the choice of experimental observables. The present proposal coming back to PAC51 has been modified to include a more detailed discussion of these concerns. 

%
%
%

\newpage

%
%
\tableofcontents

\newpage
                                                                                
%
%

\pagestyle{plain}

\pagenumbering{arabic}

\setcounter{page}{1}

%
%
%
%
%

\section{Introduction}

The challenge of the understanding of the structure and dynamics of the nucleon remains a major goal of modern Nuclear Physics despite extensive experimental scrutiny. From the initial measurements of elastic electromagnetic form factors to the accurate determination of parton distributions through deep inelastic scattering, the experiments have increased in statistical and systematic precision thanks to the development of performant electron beams together with capable detector systems. The availability of high intensity continuous polarized electron beams with high energy is providing today an unprecedented but still limited insight into the nucleon structure problem.

The Generalized Parton Distribution (GPD) para\-digm~\cite{Mul94} offers a universal and most powerful way to characterize the nucleon structure, generalizing and unifying the special cases of form factors and parton distribution functions (see~\cite{Die03,Bel05} for a review). The GPDs are the Wigner quantum phase space distribution of partons in the nucleon, describing the simultaneous distribution of particles with respect to both the position and momentum in a quantum-mechanical system~\cite{Ji03,Bel04}. They encode the correlation between partons and consequently reveal not only the spatial and momentum densities, but also the correlation between the spatial and momentum distributions, {\it i.e.} how the spatial shape of the nucleon changes when probing partons of different momentum  fraction $x$ of the nucleon. The combination of longitudinal and transverse degrees of freedom is responsible for the richness of this framework. The second moment in $x$ of GPDs are related to form factors that allow us to quantify how the orbital motion of partons in the nucleon contributes to the nucleon spin~\cite{Ji97}, and how the parton masses and the forces on partons are distributed in the transverse space~\cite{Pol03}, a question of crucial importance for the understanding of the dynamics underlying nucleon structure, and which may provide insight into the dynamics of confinement.

The mapping of the nucleon GPDs, and the detailed understanding of the spatial quark and gluon structure of the nucleon, have been widely recognized as key objectives of Nuclear Physics of the next decades. This requires a comprehensive program, combining results of measurements of a variety of processes in $eN$ scattering with structural information obtained from theoretical studies, as well as expected results from future lattice QCD calculations. Particularly, GPDs can be accessed in the  lepto-production of real photons $lN \to lN\gamma$ through Deeply Virtual Compton Scattering (DVCS) corresponding to the scattering of a virtual photon into a real photon after interacting with a parton of the nucleon. At leading twist-2, DVCS accesses the 4 quark-helicity conserving GPDs $\{H_q,E_q,\widetilde{H}_q,\widetilde{E}_q\}$ defined for each quark flavor $q\equiv\{u,d,s...\}$. They enter the cross section with combinations depending on the polarization states of the lepton beam and of the nucleon target, and are extracted from the modulation of experimental observables in terms of the $\phi$ out-of-plane angle between the leptonic and hadronic planes. This nuclear process of interest interferes with the QED (Quantum ElectroDynamics) radiation of real photons by the incoming and outgoing leptons. The non-ambiguous extraction of GPDs from experimental data not only requires a large set of observables but also the separation of the different reaction amplitudes contributing to the $lN\gamma$ reaction. The combination of measurements with lepton beams of opposite charges is an indisputable path towards such separation~\cite{Die09}. 

The physics impact of polarized and unpolarized positron beams at the Continuous Electron Beam Accelerator Facility (CEBAF) has been assessed~\cite{Acc21} and is widely recognized~\cite{Arr22,Ach23}. A strong R\&D program is currently conducted by the Ce$^+$BAF Working Group towards the implementation of positron beams at CEBAF~\cite{Hab22,Gra23,ASy23,Kaz23,Ush23}. A new positron injector is being designed based on the PEPPo (Polarized Electrons for Polarized Positrons) technique demonstrated at the CEBAF injector~\cite{Abb16}. PEPPo provides a novel and widely accessible approach based on the production, within a tungsten target, of polarized $e^+e^-$ pairs from the circularly polarized bremsstrahlung radiation of a low energy highly polarized electron beam. As opposed to other schemes operating at GeV lepton beam energies~\cite{Sok64,Omo06,Ale08}, the operation of the PEPPo technique requires only energies above the pair-production threshold and is ideally suited for the production of continuous-wave polarized positron beams.

The formal history of this proposal starts with the Letter-of-Intent LOI12-18-004~\cite{Gra18} where the perspectives of a physics program with positron beams at Jefferson Lab (JLab) were presented to the JLab Program Advisory Committee PAC46~\cite{PACRp}. As a part of this program and in the lines of PAC46 recommendations, the proposal PR12-20-009~\cite{Bur21} focusing on the measurement of Beam Charge Asymmetries (BCAs) in the DVCS channel was submitted to the JLab PAC48~\cite{PACRp}. The proposal was Conditionally Approved C2 with some concerns about the impact of positron beams and the choice of experimental observables. Providing new information about these concerns, we are proposing here to measure the unpolarized and polarized BCAs of the lepto-production of real photons on unpolarized hydrogen with {\tt CLAS12}~\cite{Bur20}, using 10.6~GeV polarized positron and electron beams at a luminosity of 0.66$\times$10$^{35}$~cm$^{-2}\cdot$s$^{-1}$. 

This proposal is organized as follows: in Sec.~\ref{sec:dvcs} we review the physics motivations for these measurements and the BCA sensitivity to GPDs in the kinematical domain of interest at {\tt CLAS12}. The impact of positron measurements is addressed in Sec.~\ref{Sec:Imp} in terms of the extraction of the physics information. In the remaining sections we discuss the experimental configuration of BCA measurements at {\tt CLAS12}, the control of systematic effects attached to the comparison of electron and positron measurements, and the detailed beam time request. Further appendix is discussing the current plans for the upgrade of the Hall B M{\o}ller polarimeter.

%
%

%
%
%
%
%

\section{Deeply Virtual Compton Scattering}
\label{sec:dvcs}

\subsection{Separation of reaction amplitudes}

\begin{figure}[!t]
\begin{center}
\includegraphics[width=0.75\textwidth]{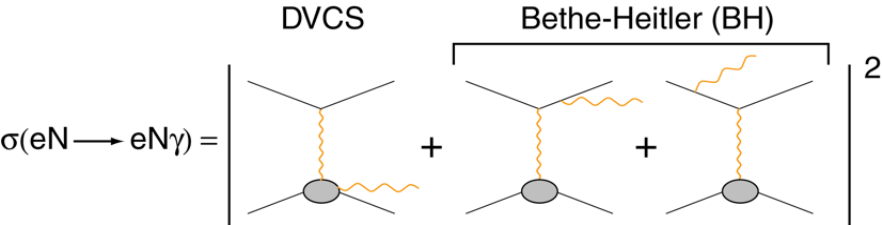}
\caption{Lowest QED-order amplitude of the electroproduction of real photons off nucleons.}
\label{EAgamma}
\end{center}
\end{figure}

Analogously to X-rays crystallography, the virtual light produced by a lepton beam scatters on the partons to reveal the details of the internal structure of the proton. For this direct access to the parton structure, the DVCS process corresponding to the reaction $\gamma^{\star}N \to \gamma N$ (Fig.~\ref{EAgamma}) is the golden channel. This process competes with the known BH reaction~\cite{Bet34} (Fig.~\ref{EAgamma}) where real photons are emitted from the initial or final leptons. The lepton beam charge ($e$) and polarization ($\lambda$) dependence of the $eN\gamma$ cross section off proton can be expressed as~\cite{Die09}
\begin{equation}
\sigma^{e}_{\lambda} = \sigma_{BH} + \sigma_{DVCS} + \lambda \, \widetilde{\sigma}_{DVCS} + e \, \left( \sigma_{INT} + \lambda \, \widetilde{\sigma}_{INT} \right) 
\end{equation}
where the index $INT$ denotes the $BH$-$DVCS$ quantum interference contribution to the cross section; ($\sigma_{BH}, \sigma_{DVCS}, \sigma_{INT}$) represent the beam polarization independent contributions of the cross section, and ($\widetilde{\sigma}_{DVCS}, \widetilde{\sigma}_{INT}$) are the beam polarization dependent contributions\footnote{($\sigma_{DVCS},\sigma_{INT}$) are related further in Sec.~\ref{Sec:BCA} to the real part of a Compton form factor, while ($\widetilde{\sigma}_{DVCS}, \widetilde{\sigma}_{INT}$) are related to its imaginary part.}. Polarized electron scattering provides the experimental observables 
\begin{eqnarray}
\sigma^-_{0} & = & \frac{\sigma^-_{+} + \sigma^-_{-}}{2} = \sigma_{BH} + \sigma_{DVCS} - \sigma_{INT} \, , \label{eq:int00} \\
\Delta \sigma^-_{\lambda} & = & \frac{\sigma^-_{+} - \sigma^-_{-}}{2} = \lambda \, \left[ \widetilde{\sigma}_{DVCS} -  \widetilde{\sigma}_{INT} \right] 
\end{eqnarray}
involving unseparated combinations of the unknown $INT$ and $DVCS$ reaction amplitudes. The comparison between polarized electron and polarized positron reactions provides the additional observables
\begin{eqnarray}
\Delta \sigma_{0}^C & = & \frac{\sigma^+_{0} - \sigma^-_{0}}{2} = \sigma_{INT} \\
\Delta \sigma_{\lambda}^C & = & \frac{\Delta\sigma^+_{\lambda} - {\Delta\sigma^-_{\lambda}} }{2} = \lambda \, \widetilde{\sigma}_{INT} 
\end{eqnarray} 
which isolate the interference amplitude. Furthermore, 
\begin{eqnarray}
\Sigma \sigma_{0}^0 & = & \frac{\sigma^+_{0} + \sigma^-_{0}}{2} = \sigma_{BH} + \sigma_{DVCS} \\
\Sigma \sigma_{\lambda}^0 & = & \frac{\Delta\sigma^+_{\lambda} + {\Delta\sigma^-_{\lambda}} }{2} = \lambda \, \widetilde{\sigma}_{DVCS} 
\end{eqnarray} 
which access a pure $DVCS$ signal. Consequently, measuring the lepto-produc\-tion of real photons off protons with polarized lepton beams of opposite charges allows to separate the four unknown contributions to the $eN\gamma$ cross section.

The essential benefit of polarized positron beams for DVCS is to provide a perfect separation of the reaction amplitudes which  consequently permits unambiguous access to GPDs. In absence of such beams, the only possible approach to this separation is to take advantage of the different beam energy sensitivity of the $DVCS$ and $INT$ amplitudes. Measurements~\cite{Def17} have shown that this Rosenbluth-like separation cannot be performed without assumptions because of higher twist and higher $\alpha_s$-order contributions to the cross section. Positron beams offer to this problem an indisputable experimental method.

\subsection{Access to Generalized Parton Distributions}

\begin{figure}[!h]
\begin{center}
\includegraphics[width=0.40\textwidth]{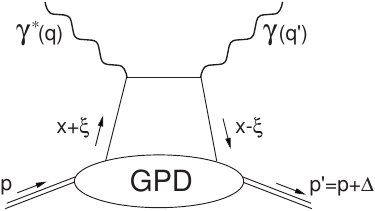}
\caption{Leading order and leading twist representation of the DVCS reaction amplitude ($+$ crossed term not shown) with main kinematic parameters of GPDs.}
\label{GPD-diag}
\end{center}
\end{figure}
GPDs are universal non-perturbative objects entering the description of hard scattering processes. Although they are not a  positive-definite probability density, GPDs correspond to the amplitude for removing a parton carrying some longitudinal momentum fraction $x$ and restoring it with a different longitudinal momentum (Fig.~\ref{GPD-diag}). The skewness $\xi \simeq x_B / (2-x_B)$, related to the Bjorken variable $x_B$=$Q^2/2M\omega$, measures the transfer of longitudinal momentum. In this process, the nucleon receives a four-momentum transfer $t={\bm{\Delta}}^2$ whose transverse component $\bm{{\Delta}_{\perp}}$ is Fourier-conjugate to the transverse distance $\bm{r_{\perp}}$ between the active parton and the center-of-mass of spectator partons in the target~\cite{Bur07}. In the limit of zero-skewness ($\xi$=$0$), GPDs can be interpreted as the Fourier transform of the distribution in the transverse plane of partons with the longitudinal momentum fraction  $x$~\cite{Bur00,Ral02,Die02,Bel02}.

GPDs enter the Leading Order (LO) $eN\gamma$ cross section through Compton Form Factors (CFF) ${\mathcal F}$ (with ${\mathcal F} \equiv \{ {\mathcal H}, {\mathcal E}, \widetilde{\mathcal H}, \widetilde{\mathcal E} \}$) defined as  
\begin{equation}
{\mathcal F}(\xi,t) = {\mathcal P} \int_{0}^{1} dx \left[ \frac{1}{x-\xi} \pm \frac{1}{x+\xi} \right] F_+(x,\xi,t) - \, i \pi \, F_+(\xi,\xi,t) \label{eq:CFF}  
\end{equation}
where ${\mathcal P}$ denotes the Cauchy's principal value integral, and  
\begin{equation}
F_+(x,\xi,t) = \sum_{q} \left(\frac{e_q}{e}\right)^2 {\left[ F^q(x,\xi,t) \mp F^q(-x,\xi,t) \right] }
\end{equation}
is the singlet GPD combination for the quark flavor $q$ where the upper sign holds for vector GPDs $(H^q,E^q)$ and the lower sign applies to axial vector GPDs  $(\tilde{H}^q,\tilde{E}^q)$. Thus the imaginary part of the CFF accesses GPDs along the diagonals $x$=$\pm\xi$ while the real part probes a convoluted integral of GPDs  over the initial longitudinal momentum of the partons. Analytical properties of the $DVCS$ amplitude lead to a dispersion relationship between the real and imaginary parts of the CFF~\cite{Ani07,Die07,Pol08}
\begin{equation}
\Re {\rm e} \left[ {\mathcal F}(\xi,t) \right] \stackrel{\rm LO}{=} D_{\mathcal F}(t) + \frac{1}{\pi}{\mathcal P}\int_{0}^{1} dx  \left
( \frac{1}{\xi-x}-\frac{1}{\xi+x}\right) \Im{\rm m} [{\mathcal F}(x,t)]
\end{equation}
where $D_{\mathcal F}(t)$ is the so-called $D$-term, a $t$-dependent subraction constant such that~\cite{Kum16}
\begin{equation}
D_{\mathcal H}(t) = - D_{\mathcal E}(t) \,\,\,\,\,\,\, D_{\widetilde{\mathcal H}}(t) = D_{\widetilde{\mathcal E}}(t) = 0 \, .
\end{equation}
Originally introduced to restore the polynomiality property of vector GPDs, the $D$-term~\cite{Pol99} enters the parameterization of the non-forward matrix element of the Energy-Momentum Tensor (EMT), which subsequently provides access to the mechanical properties of the nucleon~\cite{Pol03,Bur18,Pol18,Kum19,Bur21-1}. The independent experimental determination of the real and imaginary parts of the CFF is a key feature for the understanding of nucleon dynamics.

\subsection{Beam Charge Asymmetries} 
\label{Sec:BCA}

Considering the incident lepton $k\equiv(E,\bm{k})$ scattering into the lepton $k'\equiv(E',\bm{k'})$ at $(\theta_e,\phi_e)$ spherical angles after interaction with an unpolarized proton target at rest $p\equiv(M,\bm{0})$, the five-fold differential cross section of the $eN\gamma$ process is written
\begin{equation}
\frac{d^5\sigma_{\lambda}^e}{d^5\Omega} = \Phi_k \,  \frac{{\mathcal T}_{BH}^2 + {\mathcal T}_{DVCS}^2 + \lambda \, \widetilde{\mathcal T}_{DVCS}^2 + e \, {\mathcal T}_{INT} + e \lambda \, \widetilde{\mathcal T}_{INT}}{e^6} \label{eqXsec}
\end{equation}
where $d^5 \Omega$=$dx_B \, dQ^2 \, dt \, d\phi_e \, d \phi$ is the hypervolume subtending the elementary solid angle, and
\begin{equation}
\Phi_k = \frac{\alpha^3}{16 \pi^2} \, \frac{x_B \, y^2}{Q^4 \sqrt{1+\epsilon^2}} \label{eqfac}
\end{equation}
is a phase-space factor. The kinematical quantities in Eq.~(\ref{eqfac}) are: $y$=$p \cdot q / p \cdot k$ and $\epsilon$=$2 x_B M / Q$; $q\equiv(\omega,\bm{q})$=$k-k'$ designates the exchanged virtual photon of squared four-momentum $Q^2$=${\bm{q}}^2-\omega^2$; additionally, $p'\equiv(E_{p'},\bm{p'})$ denotes the recoil proton and $q'$=$q$+$p$-$p'$ represents the final state real photon. The different reaction amplitudes in Eq.~(\ref{eqXsec}) can be expressed as a sum of Fourier harmonics~\cite{Bel02-1} in terms of the out-of-plane angle $\phi$ between the leptonic $(\bm{k},\bm{k'})$ and hadronic $(\bm{p'},\bm{q'})$ planes, namely    
\begin{eqnarray}
{\mathcal T}_{BH}^2 & \equiv & \frac{1}{\Phi_k} \, \frac{d^5 \sigma_{BH}}{d^5 \Omega} = \frac{e^6 (1+\epsilon^2)^{-2}}{x_B^2 y^2 t \, {\mathcal P}_{1}(\phi) {\mathcal P}_{2}(\phi)} \, \sum_{n=0}^{2} c_n^{BH} \cos(n\phi) \\
{\mathcal T}_{DVCS}^2 & \equiv & \frac{1}{\Phi_k} \, \frac{d^5 \sigma_{DVCS}}{d^5 \Omega} = \frac{e^6}{y^2Q^2} \sum_{n=0}^{2} c_n^{DVCS}\cos{(n\phi)}  \\
\widetilde{\mathcal T}_{DVCS}^2 & \equiv & \frac{1}{\Phi_k} \, \frac{d^5 \widetilde{\sigma}_{DVCS}}{d^5 \Omega} = \frac{e^6}{y^2Q^2} \sum_{n=1}^{2} s_n^{DVCS}\sin{(n\phi)} \\
{\mathcal T}_{INT} & \equiv & \frac{1}{\Phi_k} \, \frac{d^5 \sigma_{INT}}{d^5 \Omega} = \frac{e^6}{x_B y^3 t \,{\mathcal P}_1(\phi) {\mathcal P}_2(\phi)} \sum_{n=0}^{3} c_n^{INT}\cos{(n\phi)} \\
\widetilde{\mathcal T}_{INT} & \equiv & \frac{1}{\Phi_k} \, \frac{d^5 \widetilde{\sigma}_{INT}}{d^5 \Omega} = \frac{e^6}{x_B y^3 t \,{\mathcal P}_1(\phi) {\mathcal P}_2(\phi)} \, \sum_{n=1}^{3} s_n^{INT} \sin(n\phi) \, .
\end{eqnarray}
The ${\mathcal T}_{BH}^2$ amplitude is exactly calculable from the electromagnetic form factors $F_1$ and $F_2$ of the proton. All other coefficients feature specific linear or bilinear combinations of CFF~\cite{Bel02-1}. The combinations (${\mathcal C}^{DVCS},{\mathcal C}^{INT}$) entering the leading twist-2 coefficients of the $DVCS$ ($c_0^{DVCS}$) and $INT$ ($c_0^{INT},c_1^{INT},s_1^{INT}$) amplitudes are 
\begin{eqnarray}
{\mathcal C}^{DVCS} & = & 4(1-x_B) {\left( {\mathcal H} {\mathcal H}^{\star} + \widetilde{\mathcal H} \widetilde{\mathcal H}^{\star} \right)} - x_B^2 {\left( {\mathcal H} {\mathcal E}^{\star} + {\mathcal E} {\mathcal H}^{\star} + \widetilde{\mathcal H} \widetilde{\mathcal E}^{\star} + \widetilde{\mathcal E} \widetilde{\mathcal H}^{\star}\right)} \nonumber \\
& & \,\,\,\,\,\,\,\,\,\, - {\left( x_B^2 + (2-x_B)^2 \frac{t}{4M^2} \right)} {\mathcal E} \widetilde{\mathcal E}^{\star} - x_B^2 \frac{t}{4M^2} \widetilde{\mathcal E} \widetilde{\mathcal E}^{\star} \\
{\mathcal C}^{INT} & = & F_1 {\mathcal H} + \xi (F_1 + F_2) \widetilde{\mathcal H} - \frac{t}{4M^2} F_2 {\mathcal E} \, .
\end{eqnarray}
The $c_{0,1}^{INT}$($s_{1}^{INT}$) coefficients are proportional to the real(imaginary) part of the ${\mathcal C}^{INT}$ combination, and the $c_0^{DVCS}$ coefficient is proportional to the real part of the ${\mathcal C}^{DVCS}$ combination. The other harmonic coefficients correspond either to twist-3 contributions ($c_1^{DVCS}$,$s_1^{DVCS}$,$c_2^{INT}$,$s_2^{INT}$) or twist-2 ($c_2^{DVCS}$,$s_2^{DVCS}$,$c_3^{INT}$,$s_3^{INT}$) double helicity-flip gluonic GPDs. Note that this elegant relationship between twist and harmonic orders, developed in the original work of Ref.~\cite{Bel02-1}, is distorted by kinematical corrections and target-mass effects~\cite{Bel10,Bra14}. This does not impact the present discussion but  reaffirms the importance of the separation of the different reaction amplitudes to provide as elementary as possible an experimental signal to allow for unambiguous interpretation. 
 
Comparing polarized electron and positron beams, the unpolarized BCA $A^C_{UU}$ can be constructed following the expresssion 
\begin{equation}
A^C_{UU} = \frac{(d^5\sigma^+_{+} + d^5\sigma^+_-) - ( d^5\sigma^-_+ + d^5\sigma^-_-)}{d^5\sigma^+_{+} + d^5\sigma^+_- + d^5\sigma^-_+ + d^5\sigma^-_-} = \frac{d^5 \sigma_{INT}}{d^5\sigma_{BH} + d^5 \sigma_{DVCS}}
\end{equation}
which, at leading twist-2, is proportional to the $\Re{\rm e} \left[{\mathcal C}^{INT}\right]$ CFF. It constitutes a selective CFF signal which becomes distorted in the case of the non-dominance of the $BH$ amplitude with respect to the polarization insensitive $DVCS$ amplitude. Similarly, the polarized BCA $A^C_{LU}$ can be constructed as
\begin{eqnarray}
A^C_{LU} & = & \frac{(d^5\sigma^+_{+} - d^5\sigma^+_-) - ( d^5\sigma^-_+ - d^5\sigma^-_-)}{d^5\sigma^+_{+} + d^5\sigma^+_- + d^5\sigma^-_+ + d^5\sigma^-_-} = \frac{\lambda \, d^5 \widetilde{\sigma}_{INT}}{d^5 \sigma_{BH} + d^5 \sigma_{DVCS}} \label{eq:BCSA}\\
 & \neq & A^-_{LU} = \frac{d^5\sigma^-_{+} - d^5\sigma^-_-}{d^5\sigma^-_{+} + d^5\sigma^-_-} = \frac{-\lambda \, ( d^5 \widetilde{\sigma}_{INT} - d^5 \widetilde{\sigma}_{DVCS} )}{d^5 \sigma_{BH} - d^5 \sigma_{INT} + d^5 \sigma_{DVCS}} \label{eq:BSAm} \\
 & \neq & A^+_{LU} = \frac{d^5\sigma^+_{+} - d^5\sigma^+_-}{d^5\sigma^+_{+} + d^5\sigma^+_-} = \frac{ \lambda \, ( d^5 \widetilde{\sigma}_{INT} + d^5 \widetilde{\sigma}_{DVCS} )}{d^5 \sigma_{BH} + d^5 \sigma_{INT} + d^5 \sigma_{DVCS}} \label{eq:BSAp}
\end{eqnarray}
which is proportional to the $\Im{\rm m} \left[{\mathcal C}^{INT}\right]$ CFF at leading twist-2. As $A^C_{UU}$, $A^C_{LU}$ is a selective CFF signal affected by the similar distortions when the $BH$ amplitude does not dominate the unpolarized cross section. At leading twist-2 and in the $BH$-dominance hypothesis, $A^C_{LU}$ is simply opposite sign to the Beam Spin Asymmetry (BSA) $A^-_{LU}$ (Eq.~(\ref{eq:BSAm})) measured with polarized electrons, and equal to the BSA $A^+_{LU}$ (Eq.~(\ref{eq:BSAp})) measured with polarized positrons. Therefore, the comparison between $A^C_{LU}$, $A^-_{LU}$, and $A^+_{LU}$ provides a handle on the validity of these hypotheses. In the case of significant differences, the neutral BSA
\begin{equation}
A^0_{LU} = \frac{(d^5\sigma^+_{+} + d^5\sigma^-_+) - ( d^5\sigma^+_- + d^5\sigma^-_-)}{d^5\sigma^+_{+} + d^5\sigma^+_- + d^5\sigma^-_+ + d^5\sigma^-_-} = \frac{\lambda \, d^5 \widetilde{\sigma}_{DVCS}}{d^5 \sigma_{BH} + d^5 \sigma_{DVCS}} 
\end{equation}
allows us to distinguish which hypothesis may not be valid.

\begin{figure}[!t]
\begin{center}
\includegraphics[width=0.440\textwidth]{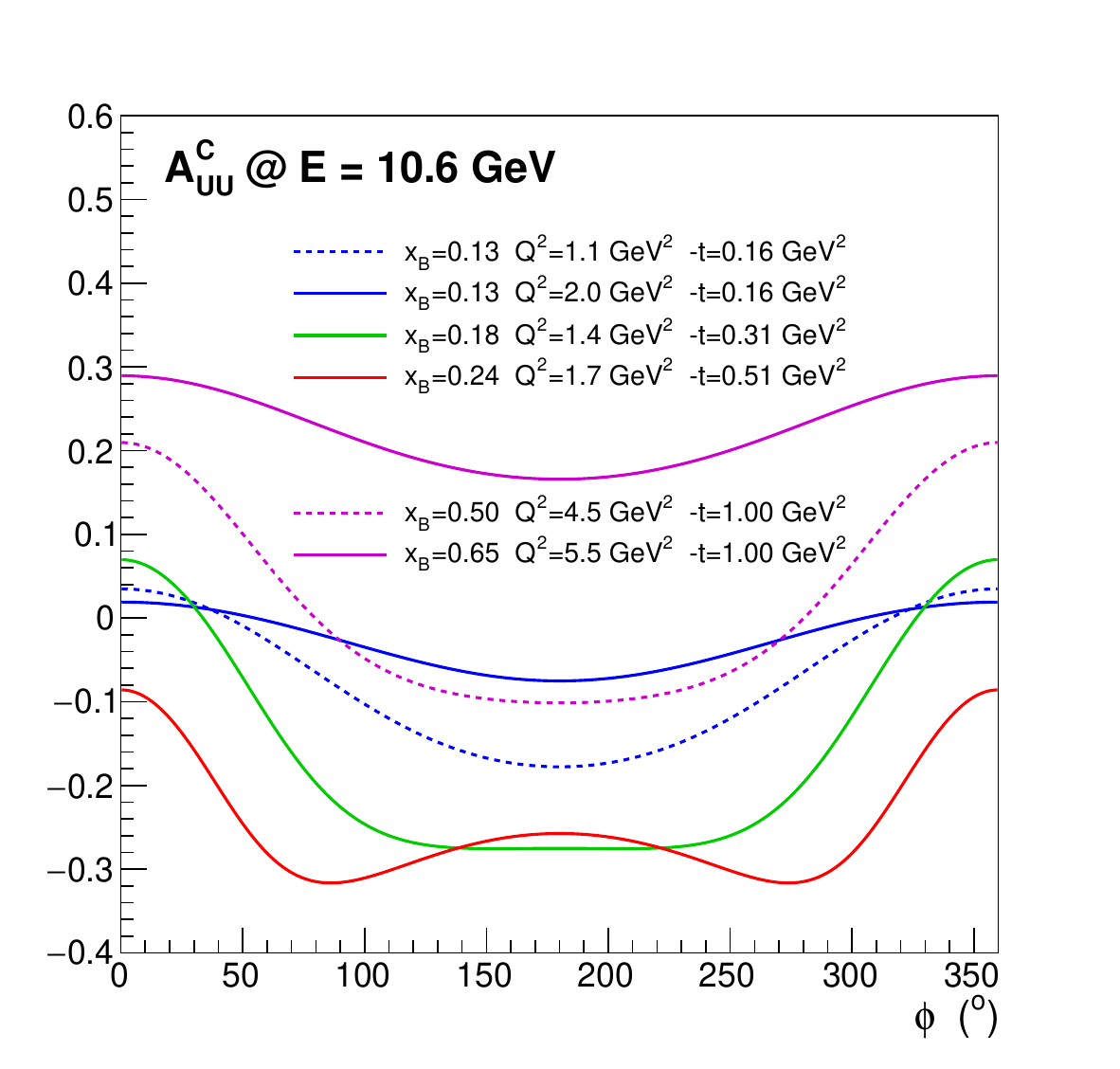}
\includegraphics[width=0.440\textwidth]{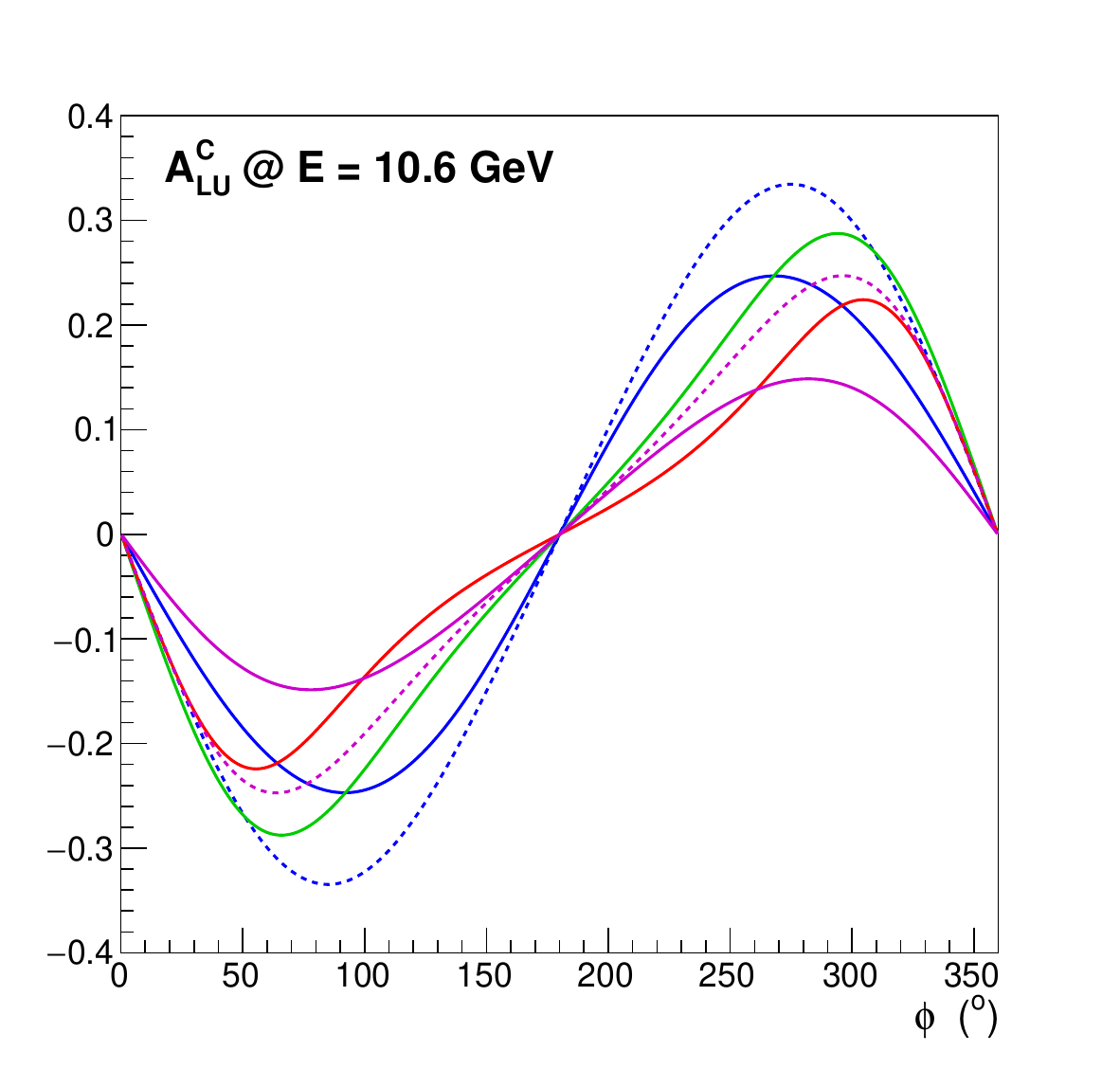}
\includegraphics[width=0.440\textwidth]{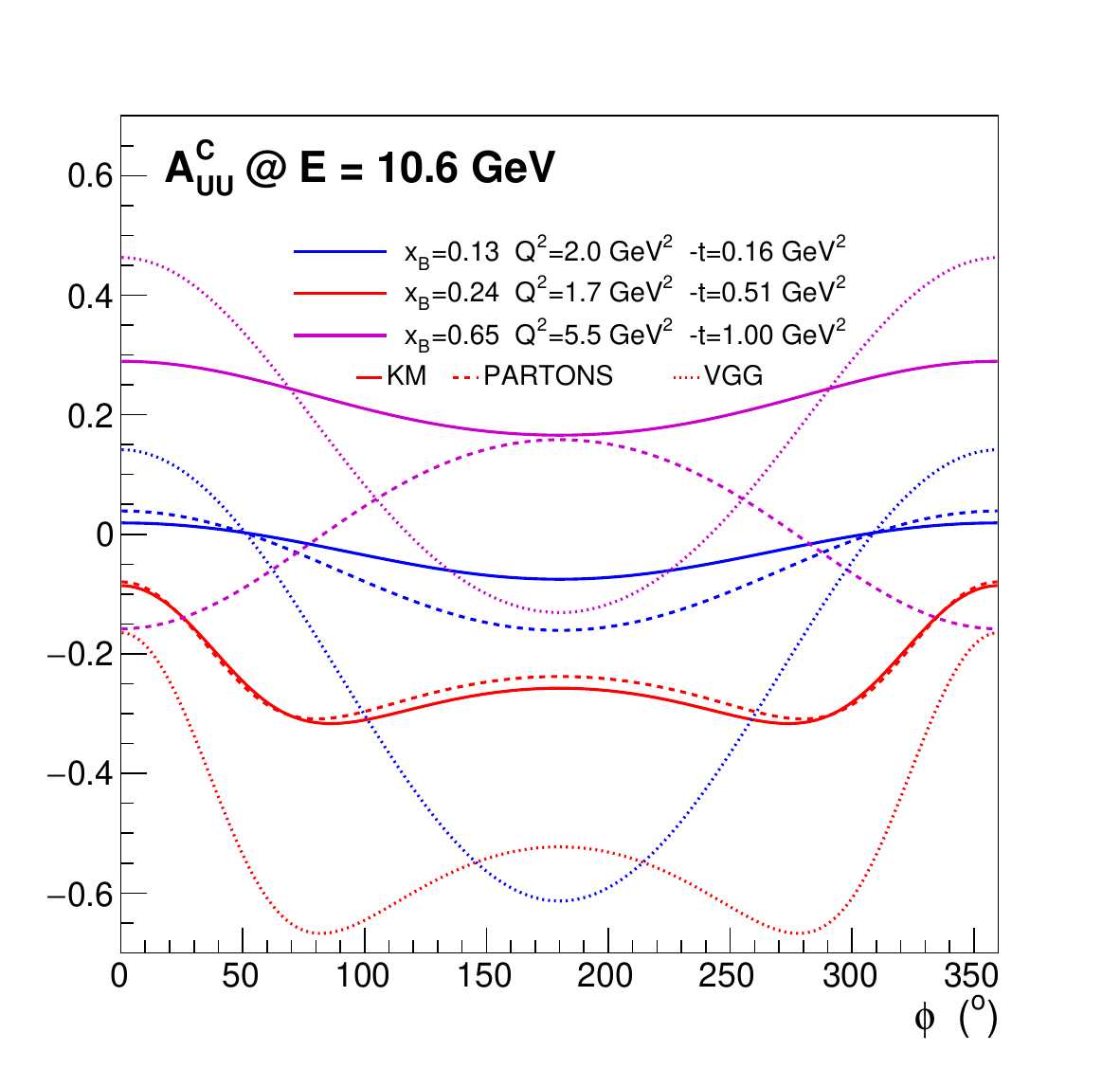}
\includegraphics[width=0.440\textwidth]{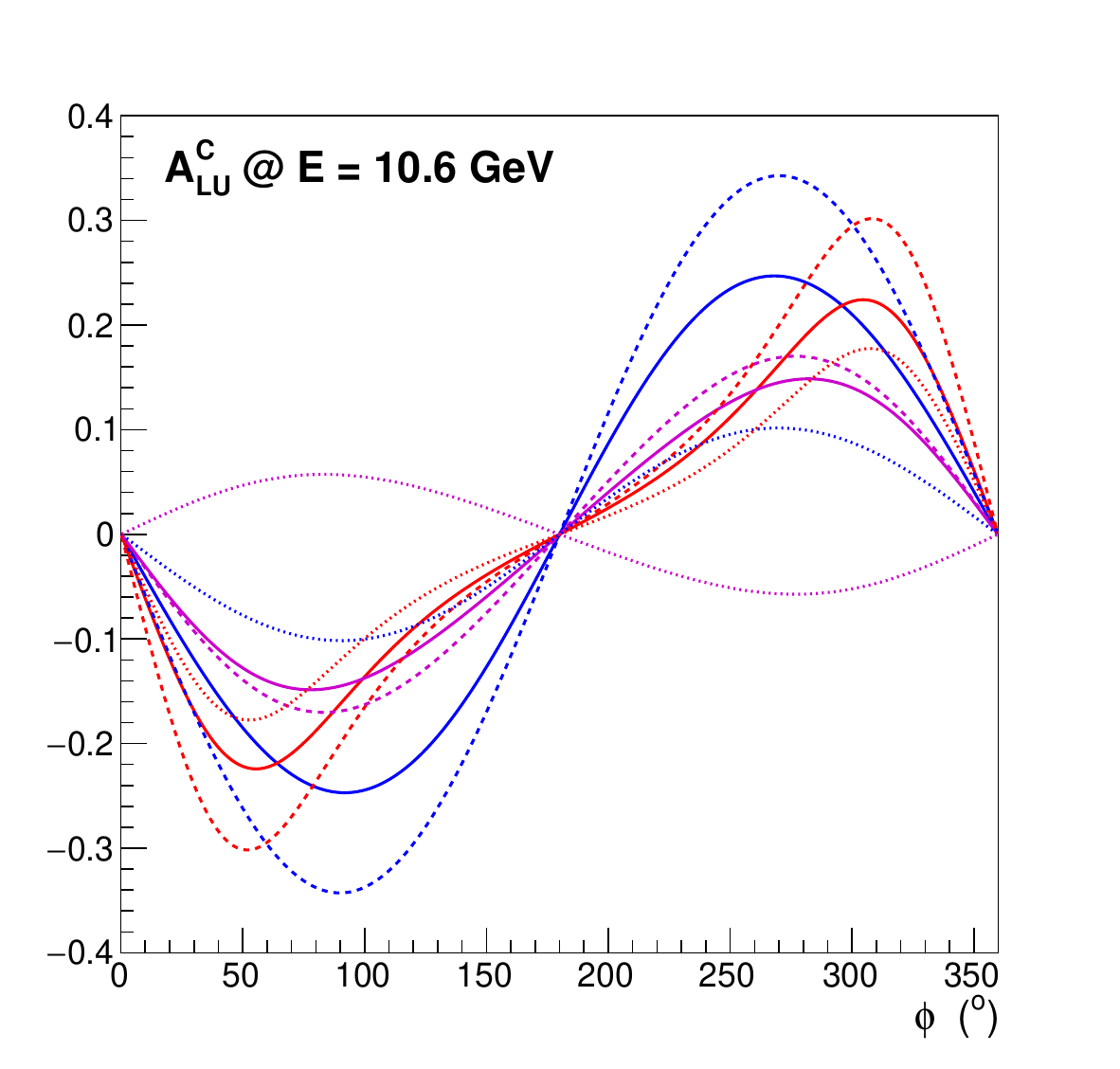}
\caption{Kinematic dependence of unpolarized (top left) and polarized (top right) BCA at a beam energy of 10.6~GeV, and CFF model sensitivity of observables (bottom panels) for selected kinematics.}
\label{BCAth}
\end{center}
\end{figure}
Unpolarized and polarized BCA observables are shown on Fig.~\ref{BCAth} (top panel) for a selected set of kinematics within {\tt CLAS12} acceptance and a 10.6~GeV beam energy. They are determined using the BM modeling of DVCS observables~\cite{Bel10} and the KM CFF~\cite{Kum10}. The sensitivity to the CFF model is also shown on Fig.~\ref{BCAth} (bottom panel) where observables calculated for 3 typical kinematics using PARTONS~\cite{Ber18} CFFs and a $\mathcal{H}$-dominated choice of VGG~\cite{Van99} CFFs are compared to previous evaluations. \newline
The amplitude of $A^C_{UU}$ strongly depends on kinematics and varies not only in magnitude (within $\pm$30\%) but also in shape, exhibiting a dominant $\cos(\phi)$ contribution eventually distorted by $\cos(2 \phi)$ contributions originating from the unpolarized part of the DVCS cross section. Similarly, the polarized BCA corresponding to the same kinematics also varies in magnitude (15\%-35\%) and shape, eventually showing a distorsion of the dominant $\sin(\phi)$ contribution by the $\cos(n \phi)$ dependence of the unpolarized DVCS cross section. A strong sensitivity of the magnitude of $A^C_{UU}$ to the CFF model is also shown (bottom panel of Fig.~\ref{BCAth}), confirming the importance of BCA observables for the extraction of the real part of the interference CFF. The dominance of the $\sin(\phi)$ modulation from the imaginary part of the interference CFF in $A^C_{LU}$ is also observed with a magnitude sensitive to the CFF model. This supports the expected {\it purity} of this observable for the

\newpage

\null\vfill

\begin{figure}[!h]
\begin{center}
\includegraphics[width=0.440\textwidth]{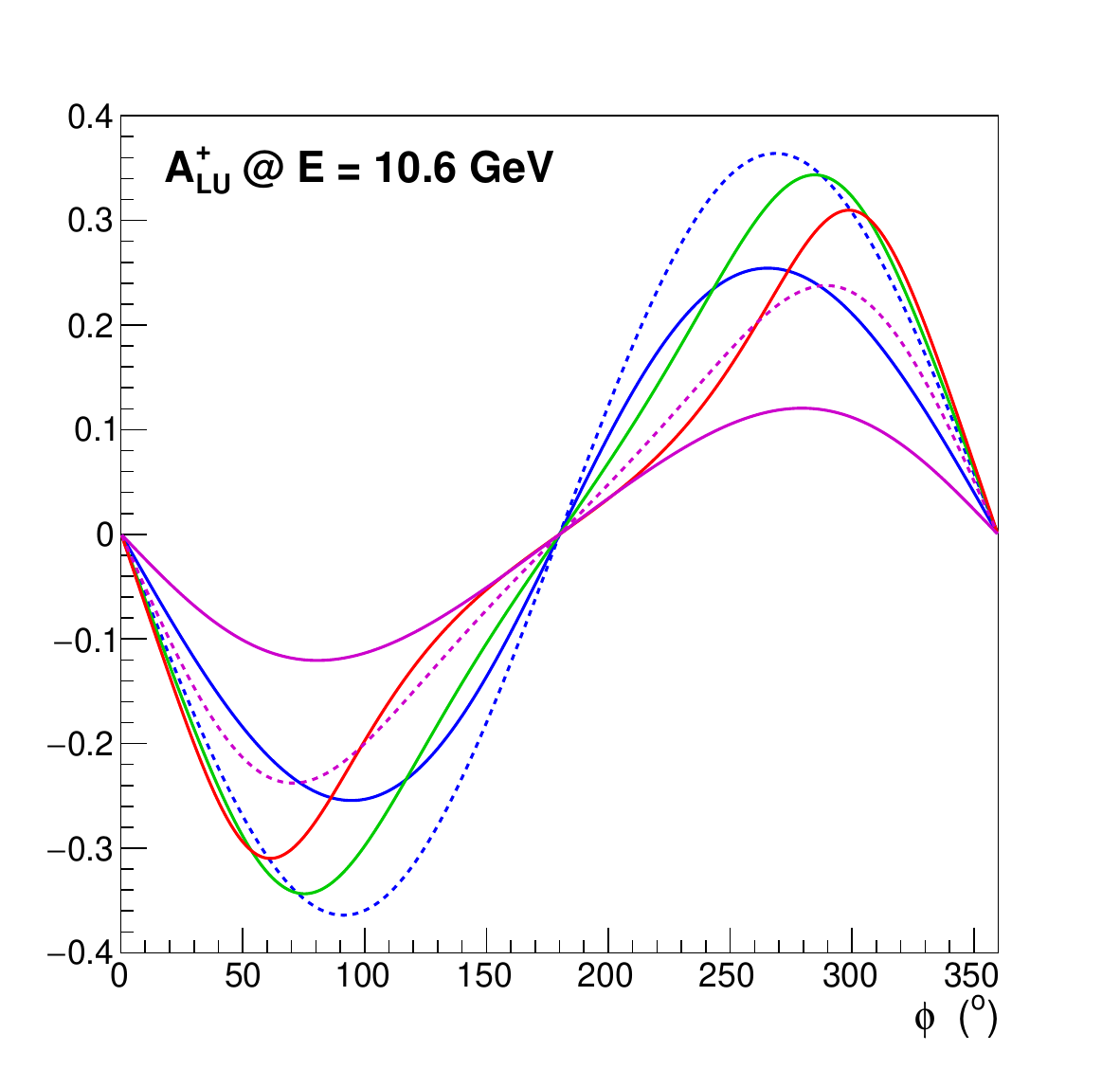}
\includegraphics[width=0.440\textwidth]{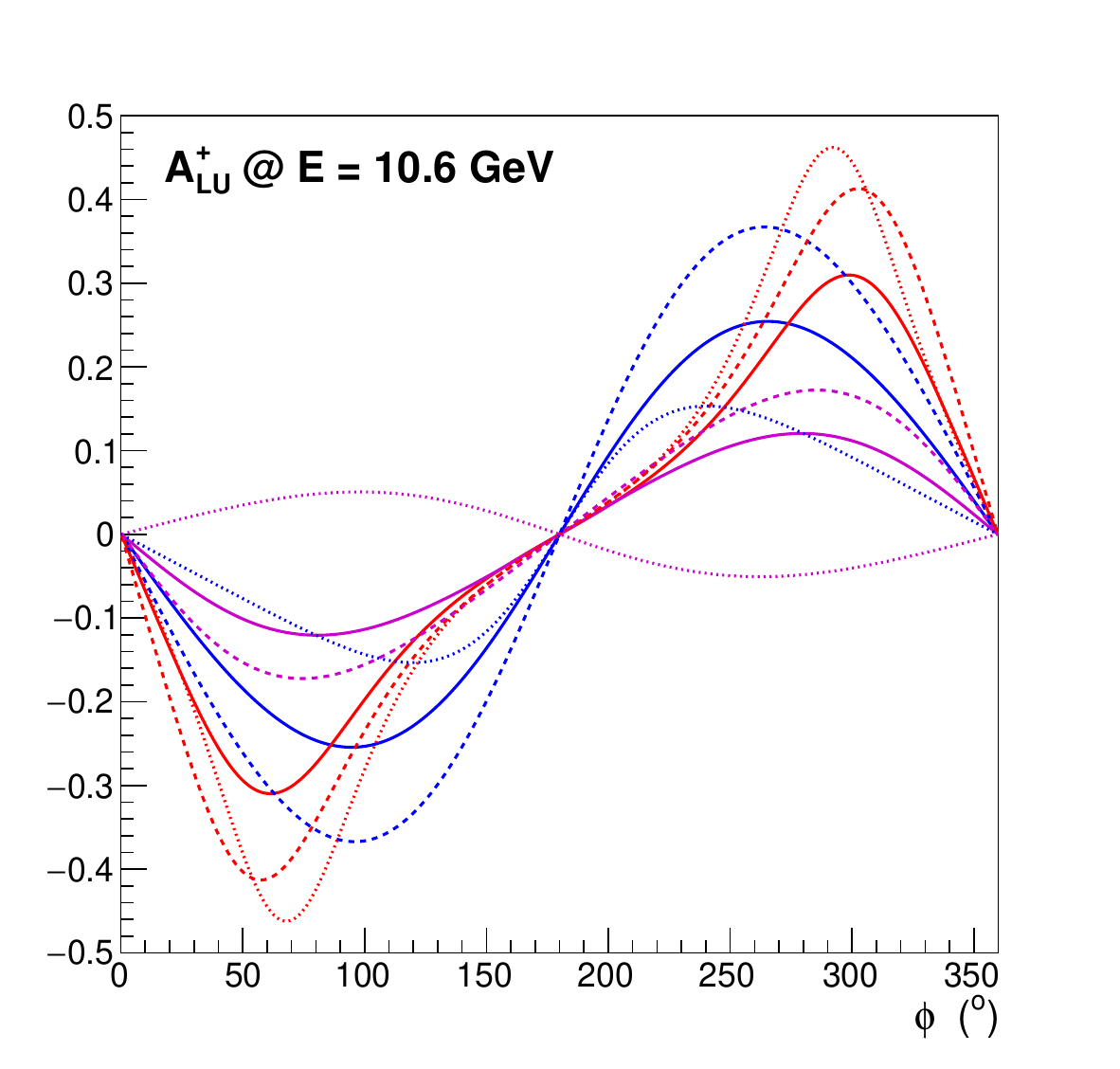}
\includegraphics[width=0.440\textwidth]{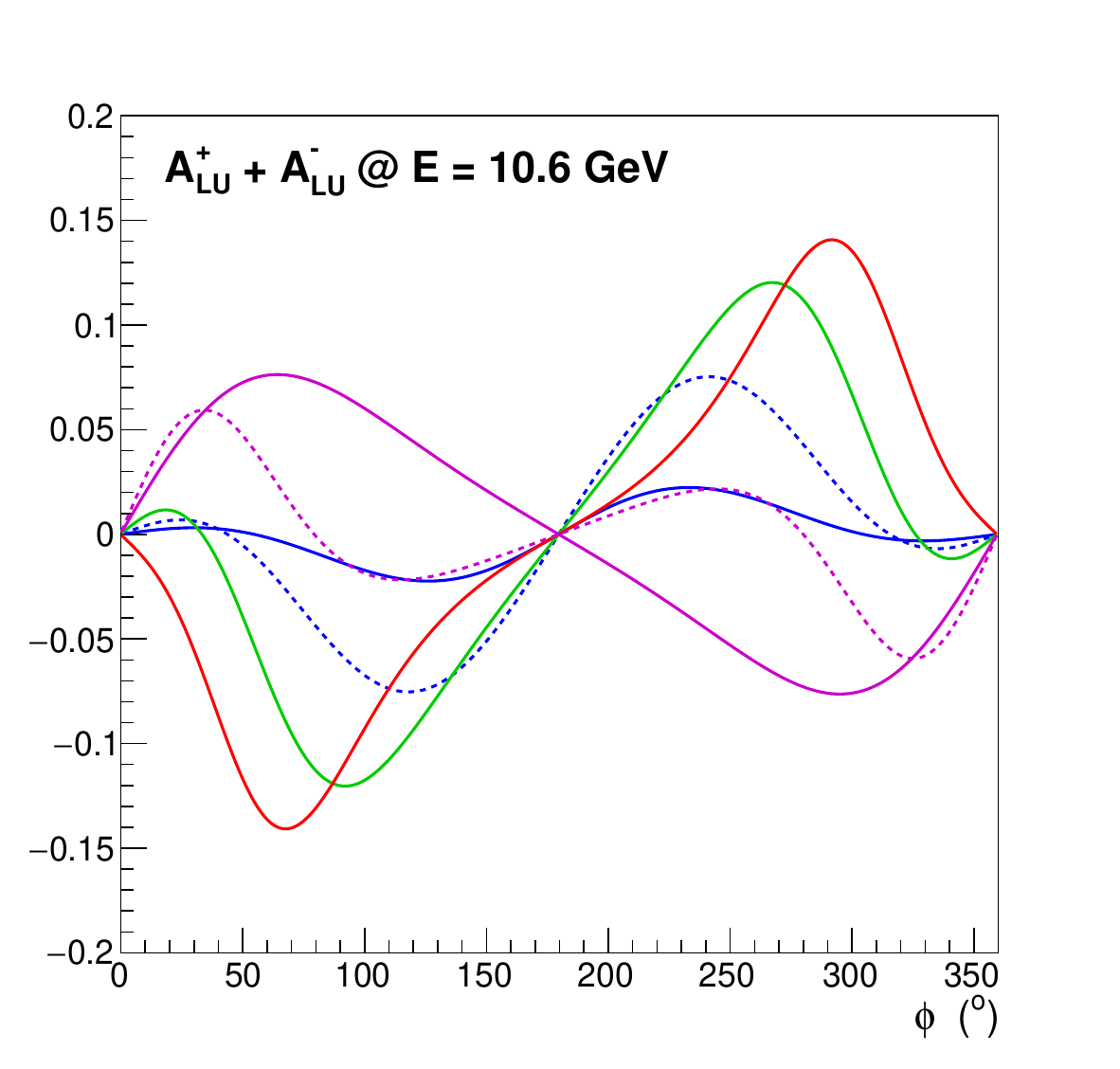}
\includegraphics[width=0.440\textwidth]{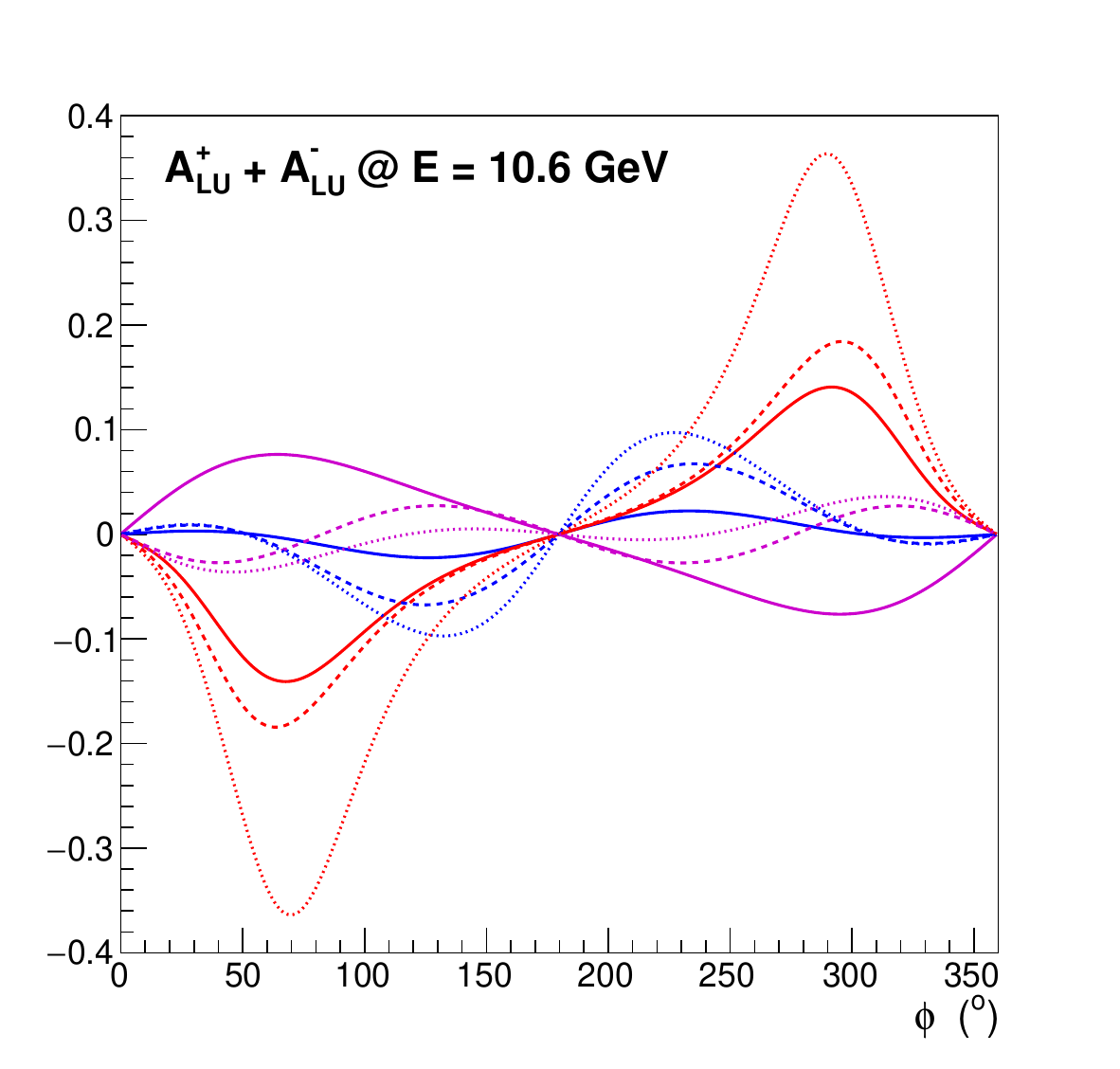}
\includegraphics[width=0.440\textwidth]{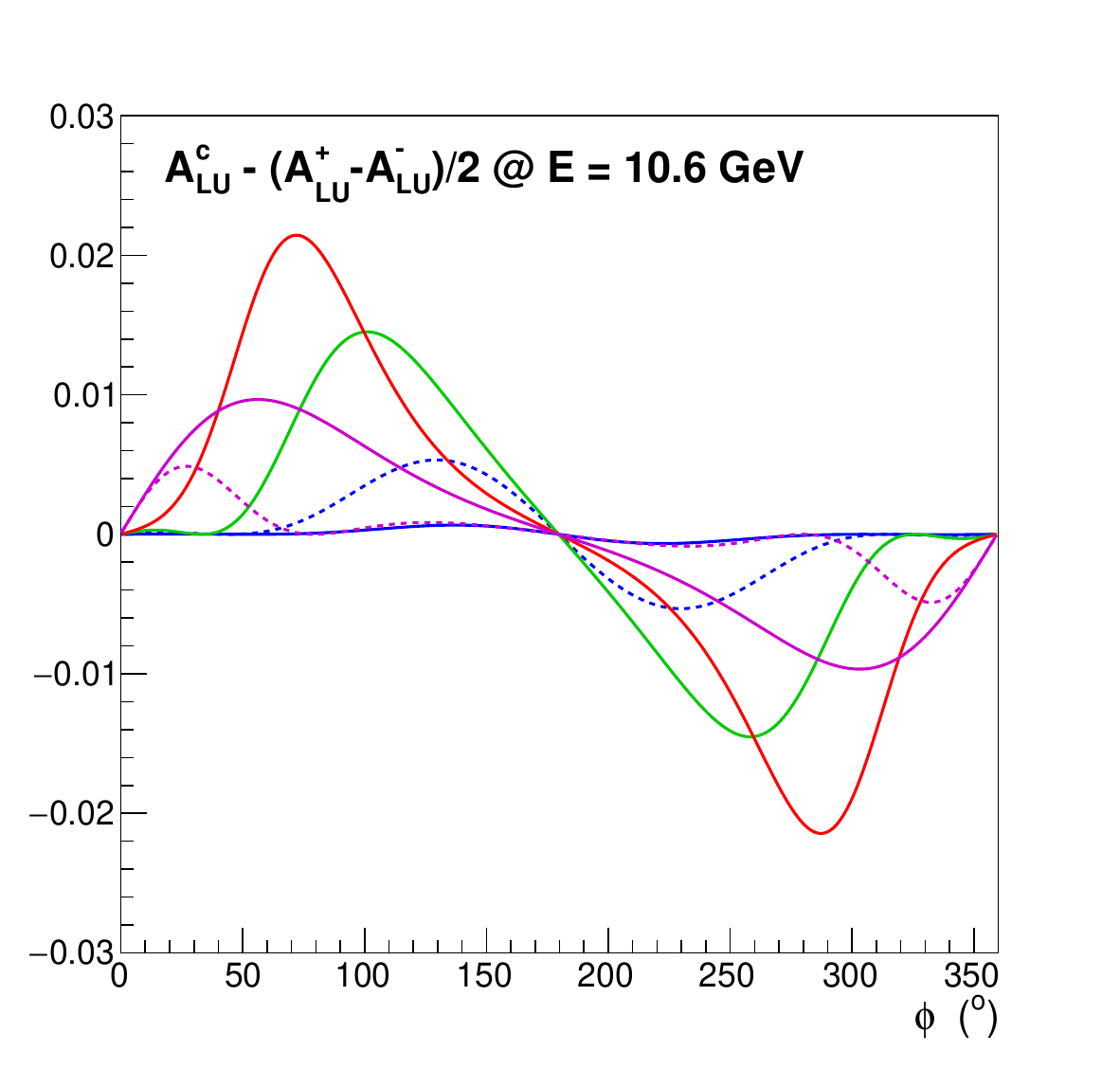}
\includegraphics[width=0.440\textwidth]{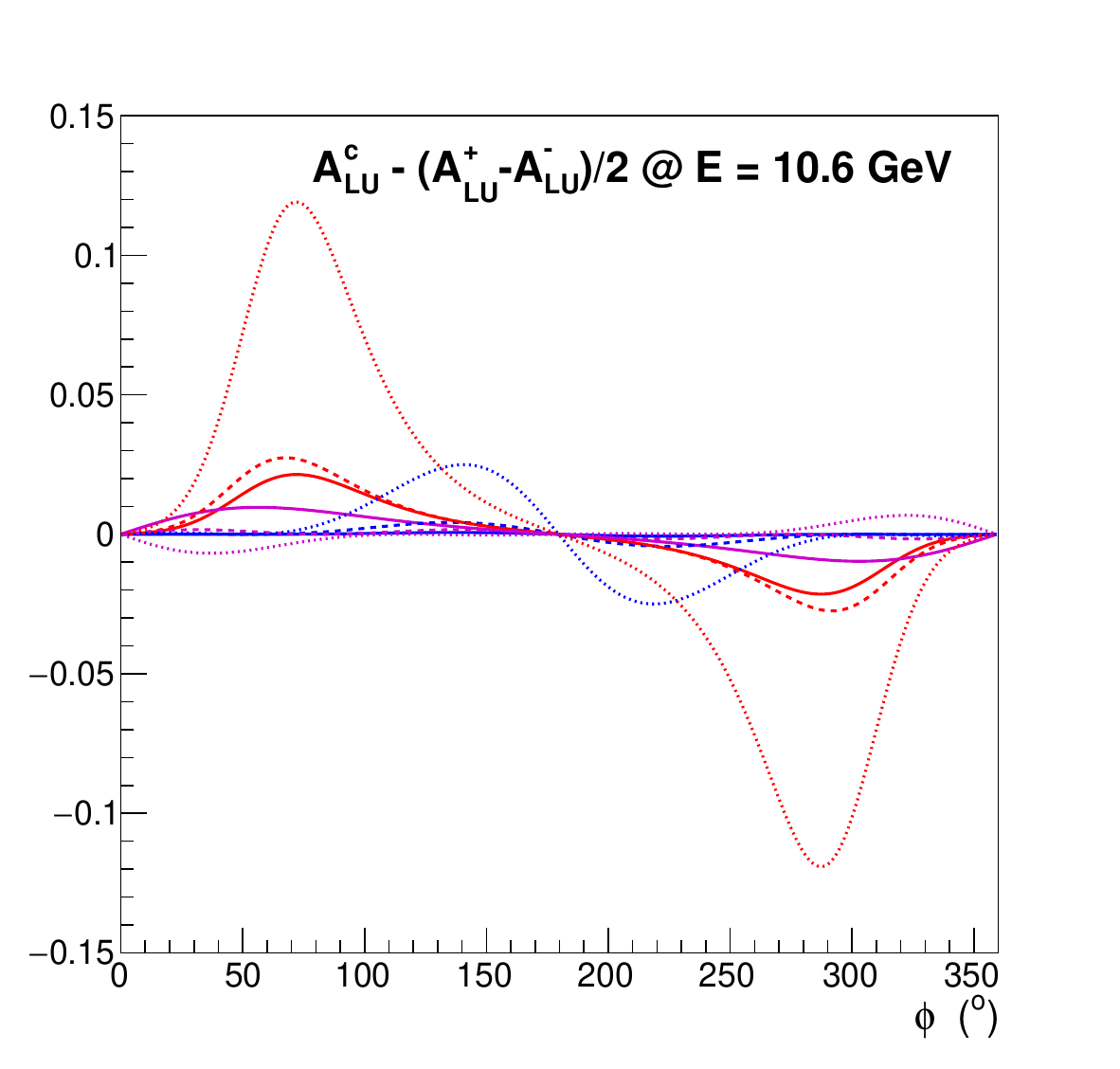}
\caption{Kinematic dependence of the positron BSA $A^+_{LU}$ (top left), the comparison $A^+_{LU}$+$A^-_{LU}$ between electron and positron BSA (middle left), the deviation $A^C_{LU}$-$(A^+_{LU}$-$A^-_{LU})$/2 from the BH-dominance hypothesis (bottom left) at a beam energy of 10.6~GeV, and CFF model sensitivity of observables (right panels) for selected kinematics. Curves labels are identical to Fig.~\ref{BCAth}.}
\label{BSAth}
\end{center}
\end{figure}

\vfill\eject

\noindent
extraction of the imaginary part of the interference CFF. 

Using a polarized positron beam, further observables can be extracted and compared with polarized electron beam observables. Particularly, the positron BSA, a golden experimental observable minimizing acceptance and efficiency effects, can be compared with the electron BSA and the polarized BCA. From Eq.~\ref{eq:BCSA}-\ref{eq:BSAp}
\begin{eqnarray}
\frac{A^+_{LU} + A^-_{LU}}{2} & = & \lambda \, \frac{d^5 \widetilde{\sigma}_{DVCS} \left( d^5 \sigma_{BH} + d^5 \sigma_{DVCS} \right) - d^5 \widetilde{\sigma}_{INT} d^5 \sigma_{INT}}{\left( d^5 \sigma_{BH} + d^5 \sigma_{DVCS} \right)^2 - \left( d^5 \sigma_{INT} \right)^2} \\
A^C_{LU} - A^+_{LU} & = & \lambda \, \frac{d^5 \widetilde{\sigma}_{INT} d^5 \sigma_{INT} - d^5 \widetilde{\sigma}_{DVCS} \left( d^5 \sigma_{BH} + d^5 \sigma_{DVCS} \right)}{\left( d^5 \sigma_{BH} + d^5 \sigma_{DVCS} \right) \left( d^5 \sigma_{BH} + d^5 \sigma_{INT} +d^5 \sigma_{DVCS} \right)} 
\end{eqnarray}
which, within the BH-dominance hypothesis, provides the relationship
\begin{equation}
A^C_{LU} = \frac{A^+_{LU} - A^-_{LU}}{2} \, .
\end{equation}

Fig.~\ref{BSAth} shows the positron BSA (top left), the comparison between positron and electron BSA (middle left) and the deviation from the BH-dominance hypothesis (bottom left) for the previous selected set of kinematics. The magnitude of $A^+_{LU}$ strongly depends on kinematics and exhibits a dominant $\sin(\phi)$ contribution. The comparison between positron and electron BSA, expressed in terms of the BSA sum, shows the expected $\phi$-modulation in absence of higher twist contributions. Both $A^+_{LU}$ and the BSA sum are strongly sensitive to the CFF model (right panel of Fig.~\ref{BSAth}). The deviations from the BH-dominance hypothesis are generally small but may become sizeable depending on the CFF scenario.   
 
%
%

%
%
%
%
%

\section{Impact of positron measurements}
\label{Sec:Imp}

The importance of BCA observables for the extraction of CFF has been stressed numerous times (see among others~\cite{Bel02-1,Die03,Bel05,Dut21}). In that  respect, HERMES~\cite{Air08,Air09,Air12} proved to be very important for the phenomenology of GPDs at moderate $x_B$~\cite{Mou18,Mou19}. Indeed, at leading twist-2 this problem can be seen as the determination of 8 unknown quantities (4$\times \Re{\rm e} \left[{\mathcal F}\right]$ and 4$\times \Im{\rm m} \left[{\mathcal F}\right]$) from a non-linear system of coupled equations which requires at a minimum 8 independent experimental observables with different sensitivities to the unknwon quantities. Dispersion relations and sum rules bring correlations between CFF and links with elastic and deep inelastic experimental data, but the problem is generally complex and requires a large set of experimental observables. Not to forget that higher twist effects and NLO corrections render the extraction even more difficult. Nevertheless, existing data about DVCS with lepton beams of opposite charges are limited and restricted to the H1~\cite{Aar09}, HERMES~\cite{Air08,Air09,Air12} and COMPASS~\cite{Akh19} experiments which explored a kinematical domain very different from the valence domain accessible at JLab.

The extraction of CFFs from DVCS observables may be classified in two generic groups: local fits where CFFs are extracted at each kinematics in a GPD model indepedent approach, and global fits which encompass a large set of experimental data at different kinematics and are sensitive to the technique or the model used to link these data~\cite{Kum16}. Both methods still depends on the cross section model (leading twist, target mass corrections, higher twist, NLO corrections...) and of further fitting hypotheses like the number of CFF to be extracted from data. In that sense, the quantitative evaluations of the impact of positron measurements discussed in this section are necessarily model dependent. The studies developed here-after are an attempt to evaluate the benefit of unpolarized and polarized BCA measurements off an unpolarized hydrogen target. This evaluation is quantified with respect to the CFF extraction performed using local or global fits of a set of experimental observables.

\subsection{Local approach evaluation}

The data sets considered in the following consists of already approved CLAS12 p-DVCS measurements with or without BCA data, considering the proposal parameters of p-DVCS experiments using a polarized electron beam with unpolarized and longitudinally polarized proton targets. Without impact on the extraction of $\mathcal{H}$, the transversely polarized target is not considered in these evaluations.  

\begin{table}[h!]
\begin{center}
\begin{tabular}{@{}r||c||c|c|c|c|c@{}} \hline \hline
                                               Observable & $\sigma_{UU}$ & $A_{LU}$ & $A_{UL}$ &   $A_{LL}$ & $A^C_{UU}$ & $A^C_{LU}$ \\ \hline
                                                 Time (d) &            80 &       80 &      100 &        100 &         80 &         80 \\ \hline
$\mathcal{L}$ ($\times$10$^{35}$ cm$^{-2} \cdot$s$^{-1}$) &           0.6 &      0.6 &        2 &          2 &        0.66 &        0.66 \\ 
                                        Hydrogen fraction &             1 &        1 &     0.17 &       0.17 &          1 &          1 \\ \hline
                                          Systematics (\%) &             5 &        3 &        3 & 3$\oplus$3 &          3 &          3 \\ \hline\hline\hline
                                                 Time (d) &            50 &       50 &       40 &         40 &         80 &         80 \\ \hline
$\mathcal{L}$ ($\times$10$^{35}$ cm$^{-2} \cdot$s$^{-1}$) &           0.6 &      0.6 &        2 &          2 &       0.66 &       0.66 \\ 
                                        Hydrogen fraction &             1 &        1 &     0.17 &       0.17 &          1 &          1 \\ \hline
                                          Systematics (\%) &            10 &        5 &        5 & 5$\oplus$5 &          5 &          5 \\ \hline\hline
\end{tabular}
\caption{Parameters of simulated observables. The upper part of the table features the parameters of the already approved {\tt CLAS12} DVCS proposals with an electron beam, referred in the text as optimum experimental conditions. The bottom part of the table shows the expected same parameters after the completion of the data taking, also referred in the text as realistic experimental conditions. The main differences concern achieved statistical and systematic uncertainties.}
\label{ObsImp}
\end{center}
\end{table}

Within a first approach, observables are determined for a 10.6~GeV beam energy using the BM modeling of the cross section~\cite{Bel10} combined with different CFFs~\cite{Van99,Kum10,Asc13,Ber18}. The projected statistical errors are obtained from the parameters of Tab.~\ref{ObsImp}, where the hydrogen fraction, corresponding to the fraction of free hydrogen protons in the target, results in a luminosity reduction factor. The CFFs $\mathcal{H}$ and $\widetilde{\mathcal{H}}$ are then simultaneously extracted from projected data using a fitting procedure which assumes the model values for the non-fitted CFFs. Individual observables are randomly smeared with the projected statistical uncertainties, and systematically shifted with the projected systematic uncertainties before CFF fitting. These two steps combined into uncertainties on the extracted CFFs. The results of the procedure repeated 
with 4 different CFF models 

\newpage

\null\vfill

\begin{figure}[!h]
\begin{center}
\includegraphics[width=\textwidth,angle=0]{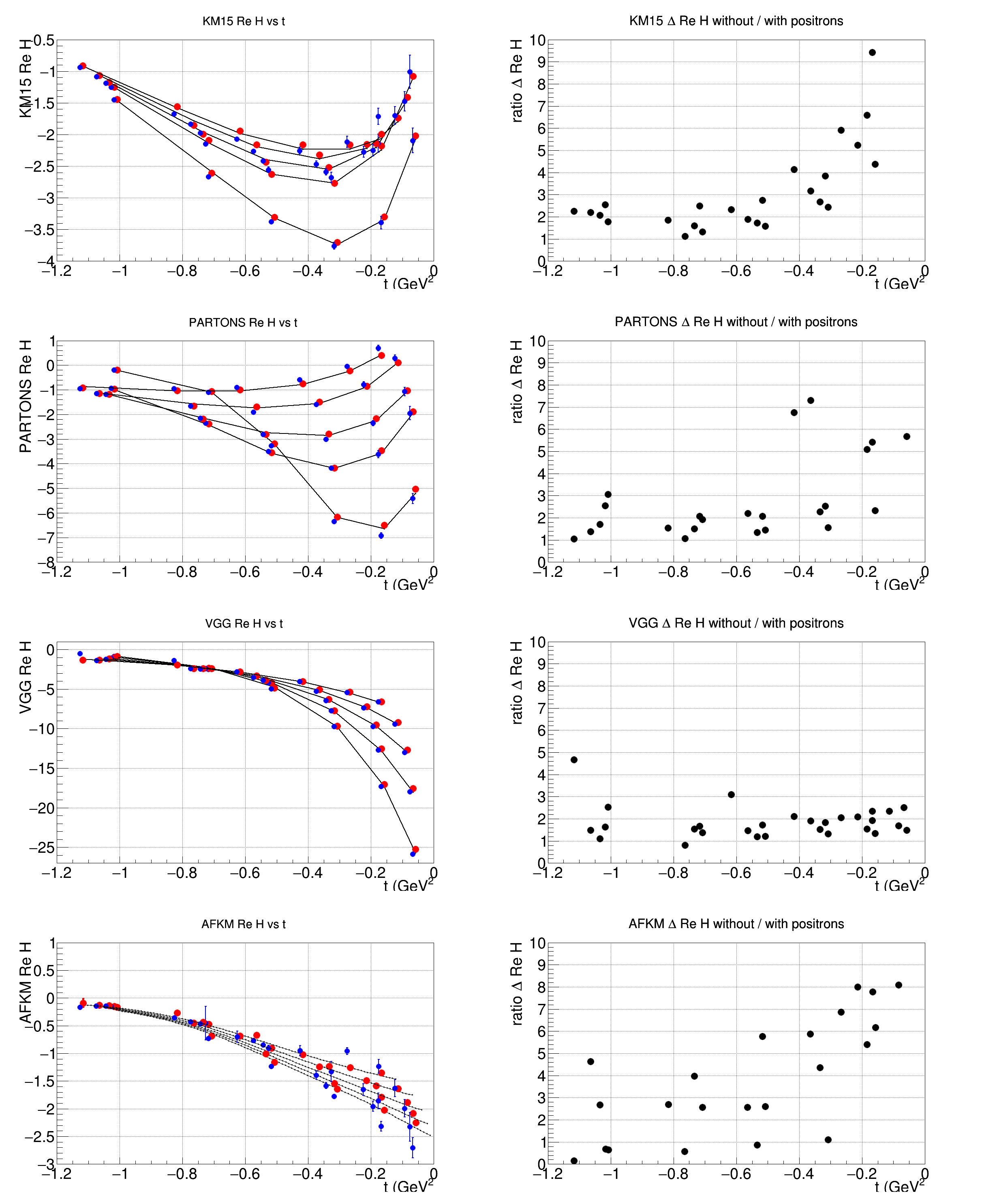}
\caption{Impact of the positron data on the extraction of $\Re {\rm e} [\mathcal{H}]$ with four different CFF model scenarios: (left panel) 
projection of extracted $\Re {\rm e} [\mathcal{H}]$ without (blue points) and with (red points) positron data compared to the model value (line); (right panel) 
ratios of errors on the extracted $\Re {\rm e} [\mathcal{H}]$ with positron data with respect to electron data only. The blue points are slightly shifted in $x$ for visual clarity.}
\label{CFF_extraction_impact}
\end{center}
\end{figure}

\vfill\eject

\null\vfill

\begin{figure}[!h]
\begin{center}
\includegraphics[width=1.00\textwidth,angle=0]{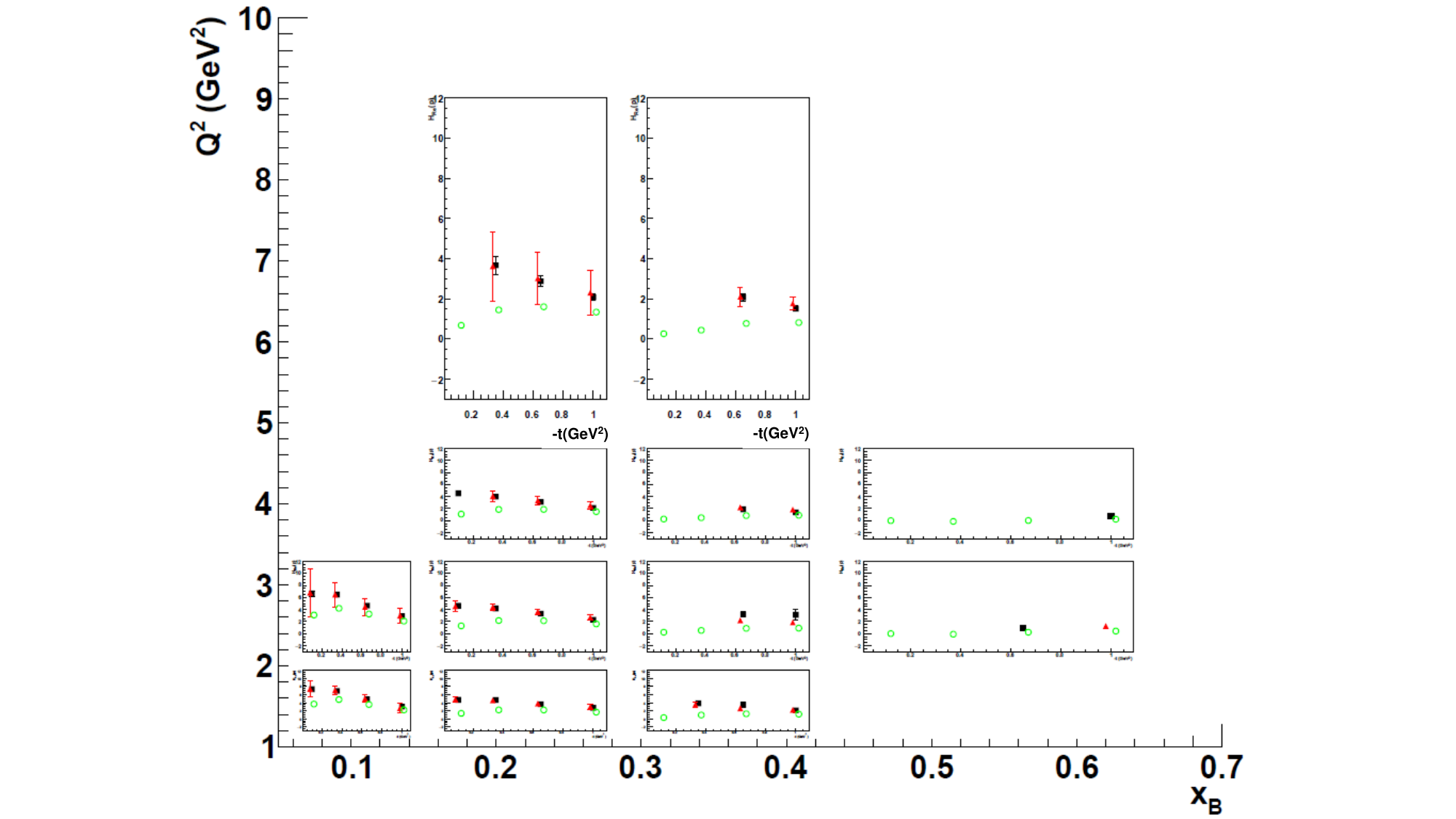}
\includegraphics[width=1.00\textwidth,angle=0]{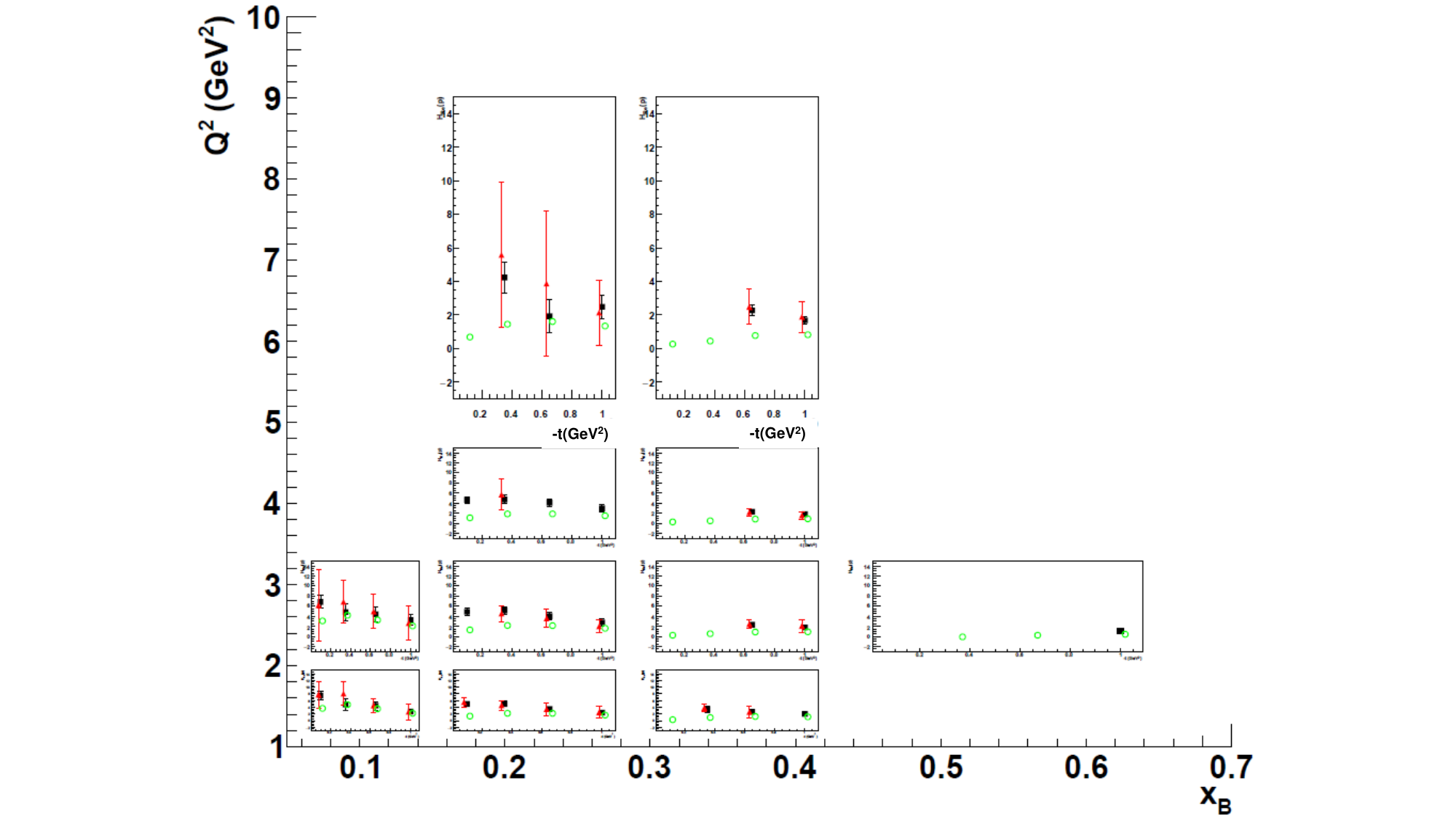}
\caption{Impact of the positron data on the extraction of $\Re {\rm e} [\mathcal{H}]$ assuming the same experimental conditions as Fig.~\ref{CFF_extraction_impact} for extracting 4 CFFs (top panel), or assuming the realistic experimental conditions of Tab.~\ref{ObsImp} for the extraction of 7 CFFs (bottom panel). The benefit of positron data can be seen from the comparison of red (without positron data) and black (with positron data) points while green points represent the model value. The red points are slightly shifted in $-t$ for visual clarity.}
\label{VGGImp}
\end{center}
\end{figure}

\vfill\eject

are summarized in Fig~\ref{CFF_extraction_impact}. The left column shows the model $\Re {\rm e} [\mathcal{H}]$ as a function of $t$ for different $(x_B,Q^2)$ bins (line), together with the extracted values without (blue points) and with (red points) the positron data. The corresponding set of plots on the right column shows the ratios of the total uncertainties, statistical and systematic errors added in quadrature. The impact of the positron data is found to be particularly strong at small $\vert t \vert$ where they can decrease uncertainties on $\Re {\rm e} [\mathcal{H}]$ by over a factor five. The electron data only scenario tends to give values different from the model values. By providing a pure interference signal, positron data corrects for this deviation and allows the fitting procedure to recover the input model value. In the case experimental observables are dominated by $\mathcal{H}$, correlations between CFF are reduced and the impact of positron data should be minimal. This is indeed observed in the projections of the model labelled VGG where the positron impact is consistent with the statistics added by positron measurements.    

Within a second approach, the same experimental observables are generated using the VGG model~\cite{Van99} and a CFF parameterization which fairly reproduces 6~GeV CLAS data. In a first study, the $\mathcal{H}$ and $\widetilde{\mathcal{H}}$ CFFs are extracted for the same set of observables considering optimum experimental conditions of Tab.~\ref{ObsImp}, and in a second one the full set of 7 CFFs are extracted assuming $\Im {\rm m} [\widetilde{\mathcal{E}}]$=$0$ and considering the realistic experimental conditions of Tab.~\ref{ObsImp}. In both cases, the VGG modeling of experimental observables is used for the local fitting procedure~\cite{Gui08}. The comparison of the black (with $e^+$) and red (without $e^+$) points signs the impact of positron measurements while the comparison with respect to green points tells about the ability to recover the considered model value. The observed differences in the latter comparison originate from the lack of experimental constraints on the other CFFs entering experimental observables. Both scenarios predict a strong impact of positron data, even more determinant in the realistic one where electron data only do not always succeed to extract a meaningful value for $\Re {\rm e} [\mathcal{H}]$.

\subsection{Global approach evaluation}

\begin{figure}[!h]
\begin{center}
\includegraphics[width=0.9\textwidth,angle=0]{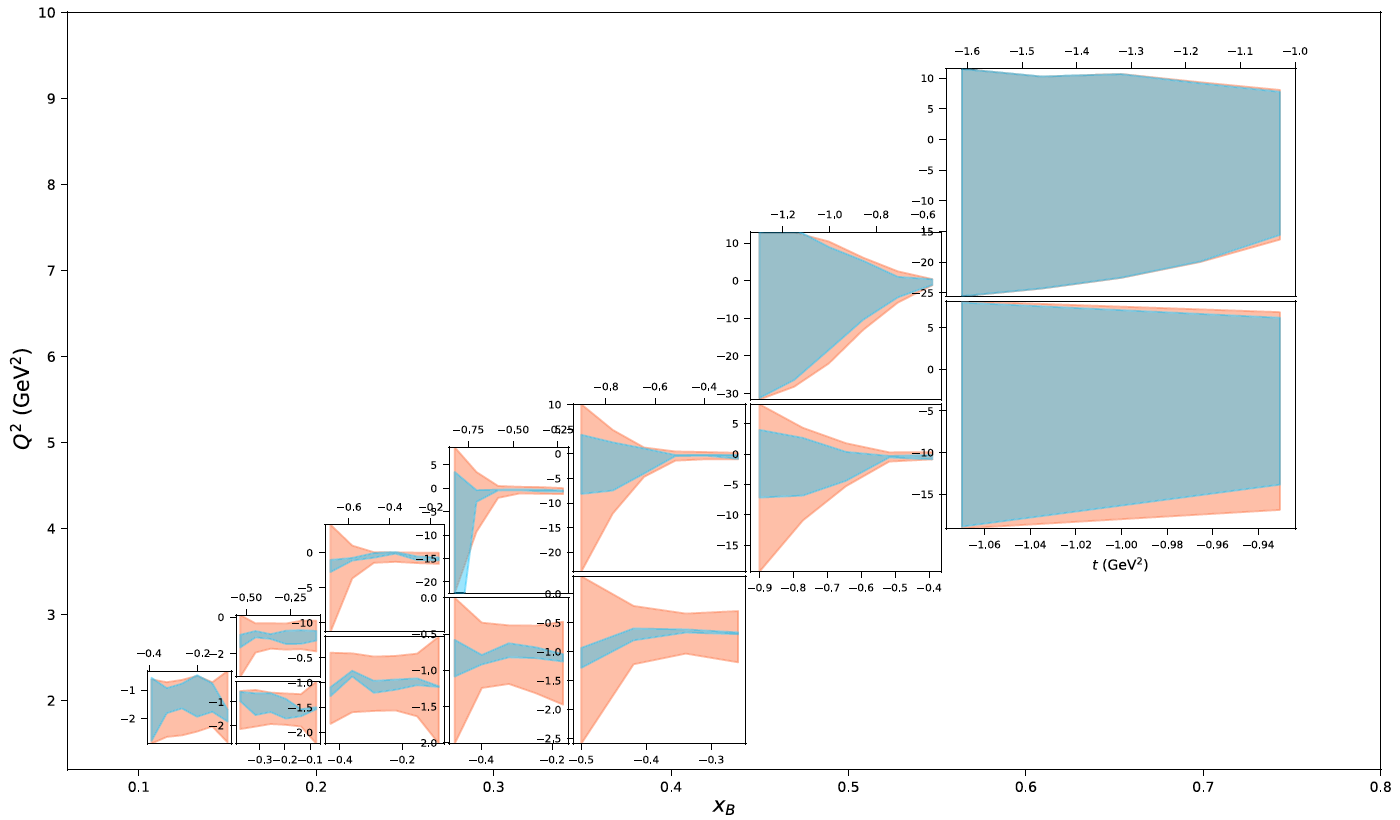}
\caption{$t$-dependence of the 68\% confidence region for $\Re{\rm e} \left[\mathcal{H}\right]$ without (red band) and with (blue band) positron observables~\cite{Dut21}.}
\label{GloImp}
\end{center}
\end{figure}
The benefit of global fits of DVCS experimental data is the capability to describe measurements from different experiments at once, whatever experimental observable or kinematics are concerned. Several techniques have been proposed to supply the link between the different kinematical domains from simple modeling of CFFs based on a parameterized GPD ansatz, up to physically-motivated parameterization of the CFFs~\cite{Mou18}, and artificial neural network (ANN) techniques~\cite{Mou19}. The impact of the measurement of beam charge asymmetries presented in Ref.~\cite{Dut21} and summarized here, is evaluated from an ANN analysis of all existing data from ZEUS, H1, HERMES, Hall A, and CLAS experiments. Projected BCA observables are computed according to this ansatz and their impact is evaluated from a Bayesian reweigthing procedure using optimum experimental conditions of Tab.~\ref{ObsImp}. The results reported in Fig.~\ref{GloImp} show the 68\% confidence level on the value of $\Re{\rm e} \left[\mathcal{H}\right]$ without (red band) and with BCAs (blue band). As observed in the local approach, the benefit of a positron beam is particularly striking at small $\vert t \vert$, a kinematical region of high interest for the interpretation of the DVCS process. It becomes marginal at large $(x_B,Q^2)$ because of lack of statistics. Nevertheless, the improvement of the quality of the $Q^2$-coverage of $\Re{\rm e} \left[\mathcal{H}\right]$ should be noted as a tool of interest for the determination of the $D$-term~\cite{Dut21-1}. 

In summary, both local and global evaluations predict a high impact of positron data with a reduction of the error bars on the value of $\Re {\rm e} [\mathcal{H}]$ by a factor 3-5 on average. Isolating interference contributions, BCAs provide a pure signal which benefits the extraction of $\Re {\rm e} [\mathcal{H}]$ by reducing the correlations between CFFs. This feature cannot be achieved using single charge lepton beams. Such an improvement is particularly relevant for the experimental determination of the mechanical properties of the proton and for universality studies of GPDs combining DVCS and Time-like Compton Scattering~\cite{Cha21}.

%
%

%
%
%
%
%

\section{p-DVCS at {\tt CLAS12} with a positron beam}

\subsection{Detector configuration}

\begin{figure}[!h]
\centering{\hspace{-0.5cm}{\includegraphics[width=0.85\columnwidth]{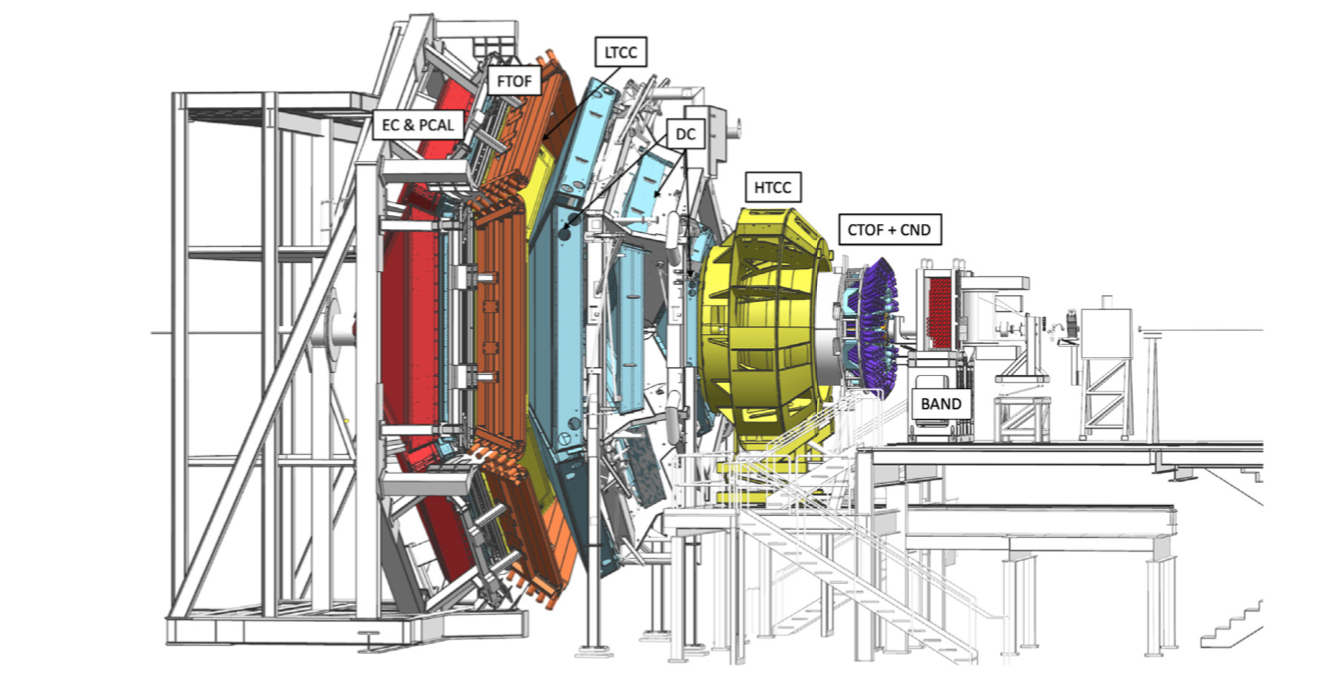}}}
\caption{{\tt CLAS12} in Hall B. The positron beam comes from the right and hits the target in the center of the solenoid magnet, which is at the core of the Central Detector (CD). It is largely hidden from view inside the HTCC \v{C}erenkov counter.} 
\label{clas12}
\end{figure}
\begin{figure}[!t]
\centering{\includegraphics[width=0.75\columnwidth]{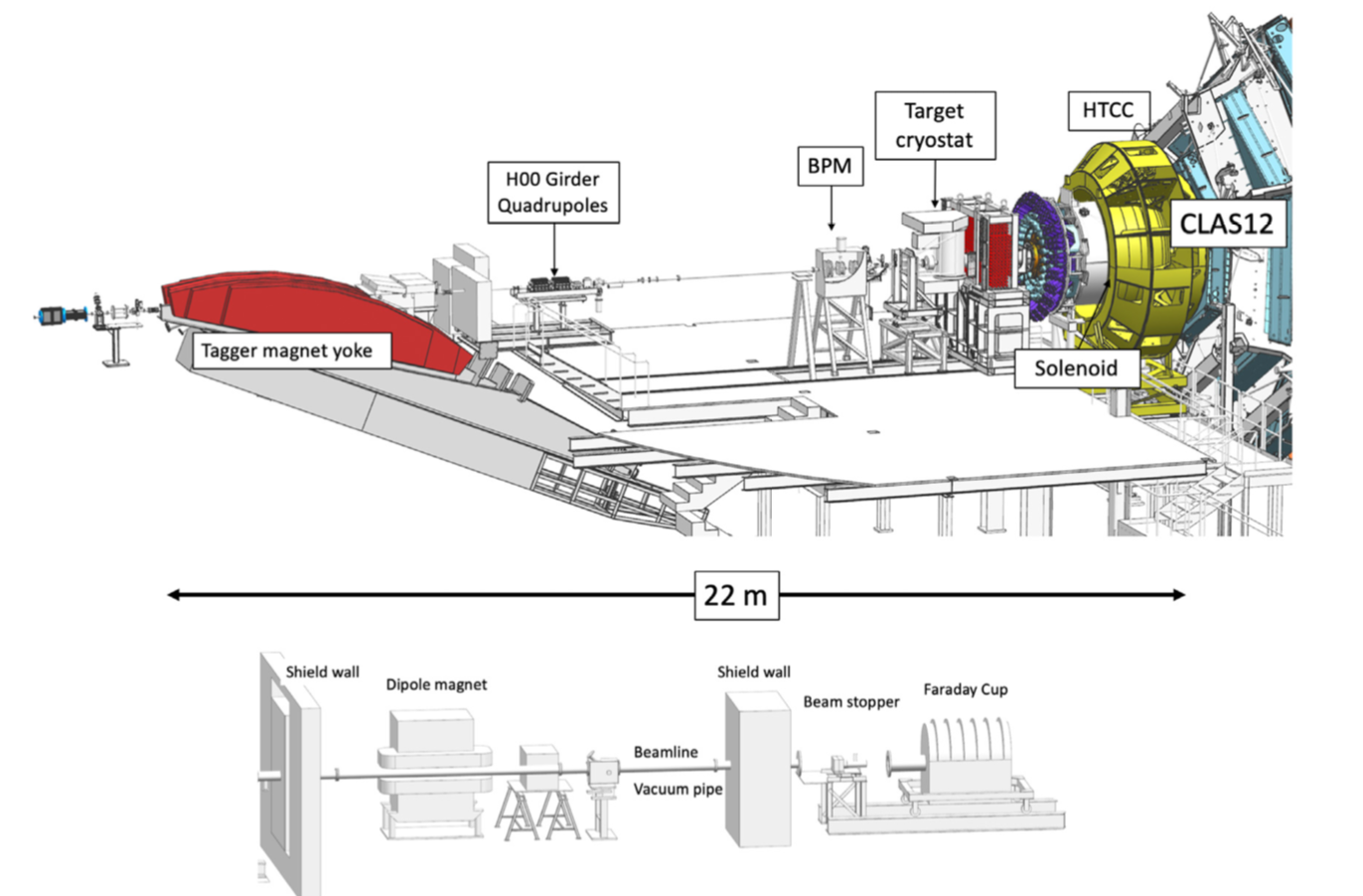}}
\caption{ Hall B beamline.} 
\label{hallb-beamline}
\end{figure}

 
The experiment will measure the DVCS process $e^+p \to e^+p\gamma$ with the {\tt CLAS12} spectrometer. The arrangement of {\tt CLAS12} in the Hall B is shown in a side view in Fig.~\ref{clas12}, and the beam line upstream and downstream of {\tt CLAS12} are shown in Fig.~\ref{hallb-beamline} (see Ref.~\cite{Bur20} for details). When operating with positron beam the experiment will use the standard Hall B beam line with the electrical diagnostics in reversed charge mode from operating the beam line and the experimental equipment with electron beam. This includes the nano-ampere beam position and current monitors, the beam line magnetic elements including the tagger magnet, which is energized during beam polarization measurements, and the charge integrating Faraday cup. The experimental setup will be identical to the standard electron beam setup with both magnets, the Solenoid and the Torus magnet in reversed current mode from electron scattering experiments. As the positron beam emittance at the target will be larger than in standard electron beam operation, the liquid hydrogen target cell will be redesigned with increased entrance and exit window sizes.
 
\subsection{Kinematic coverage}

\begin{figure}[!ht]
\begin{center}
\includegraphics[width=0.875\columnwidth]{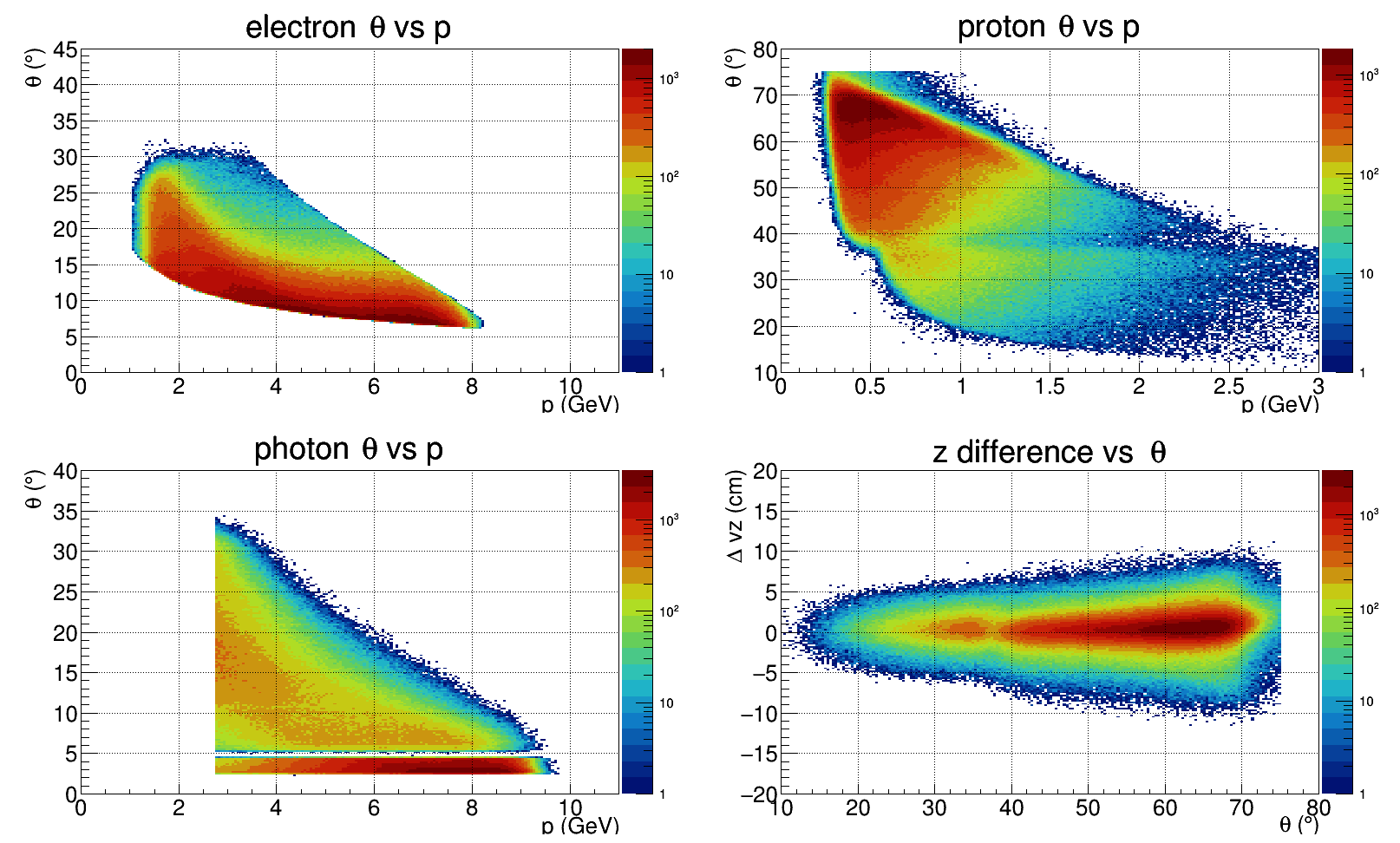}
\caption{Kinematic coverage of {\tt CLAS12} for exclusive DVCS events at a beam energy of 10.6~GeV: scattered lepton reconstruction coverage in polar angle versus momentum (top left); proton reconstruction coverage in polar angle versus momentum (top right); at polar angles close to 40$^\circ$, protons are partially reconstructed in the FD and partially in the CD; high-energy photon detection coverage in polar angle versus photon energy (bottom left); the narrow band below 5$^\circ$ indicates photon detection in the FT calorimeter; difference in reconstructed $z$-vertex for scattered leptons and the recoil protons (bottom right).}
\label{dvcs-coverage}
\end{center}
\end{figure}
The simultaneous kinematic coverage of the DVCS process in the {\tt CLAS12} acceptance is shown in Fig.~\ref{dvcs-coverage} from a subset of Run Group A (RGA) data and a detector configuration similar to the positron configuration {\it i.e.} Torus in OUT-Bending mode and FTCal ON. Scattered electrons/positrons will be detected in the {\tt CLAS12} Forward Detectors (FD) including the high threshold \v{C}erenkov Counter (HTCC), the drift chamber tracking system, the Forward Time-of-Flight system (FTOF) and the electromagnetic calorimeter (ECAL). The latter consists of the pre-shower calorimeter (PCAL) and the EC-inner and EC-outer parts of the electromagnetic calorimeter (EC) providing a 3-fold longitudinal segmentation. DVCS photons are measured in the {\tt CLAS12} ECAL that covers the polar angle range from about 5$^\circ$ to 35$^\circ$. Additionally, high energy photons are also detected in the Forward Tagger calorimeter (FTCal), which spans the polar angle range of 2.5$^\circ$ to 4.5$^\circ$. Protons are detected mostly in the {\tt CLAS12} Central Detector (CD) with momenta above 300~MeV/$c$, but a significant fraction is also detected in the {\tt CLAS12} FD, especially those in the higher $-t$ range.  
\begin{figure}[!t]
\begin{center}
\includegraphics[width=0.875\columnwidth]{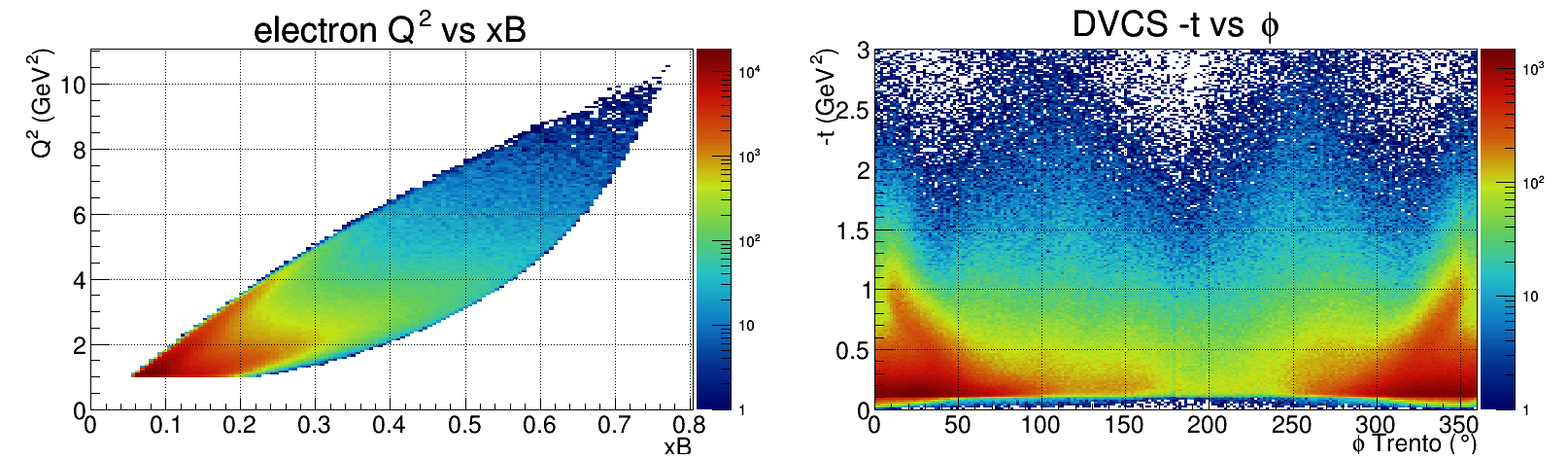}
\caption{Kinematic coverage of exclusive DVCS/BH events in $Q^2$ versus $x_B$ (left), and in $-t$ (right) plotted versus the azimuthal $\phi$-dependence.}
\label{dvcs-kine}
\end{center}
\end{figure}

\begin{figure}[!t]
\begin{center}
\includegraphics[width=0.875\columnwidth]{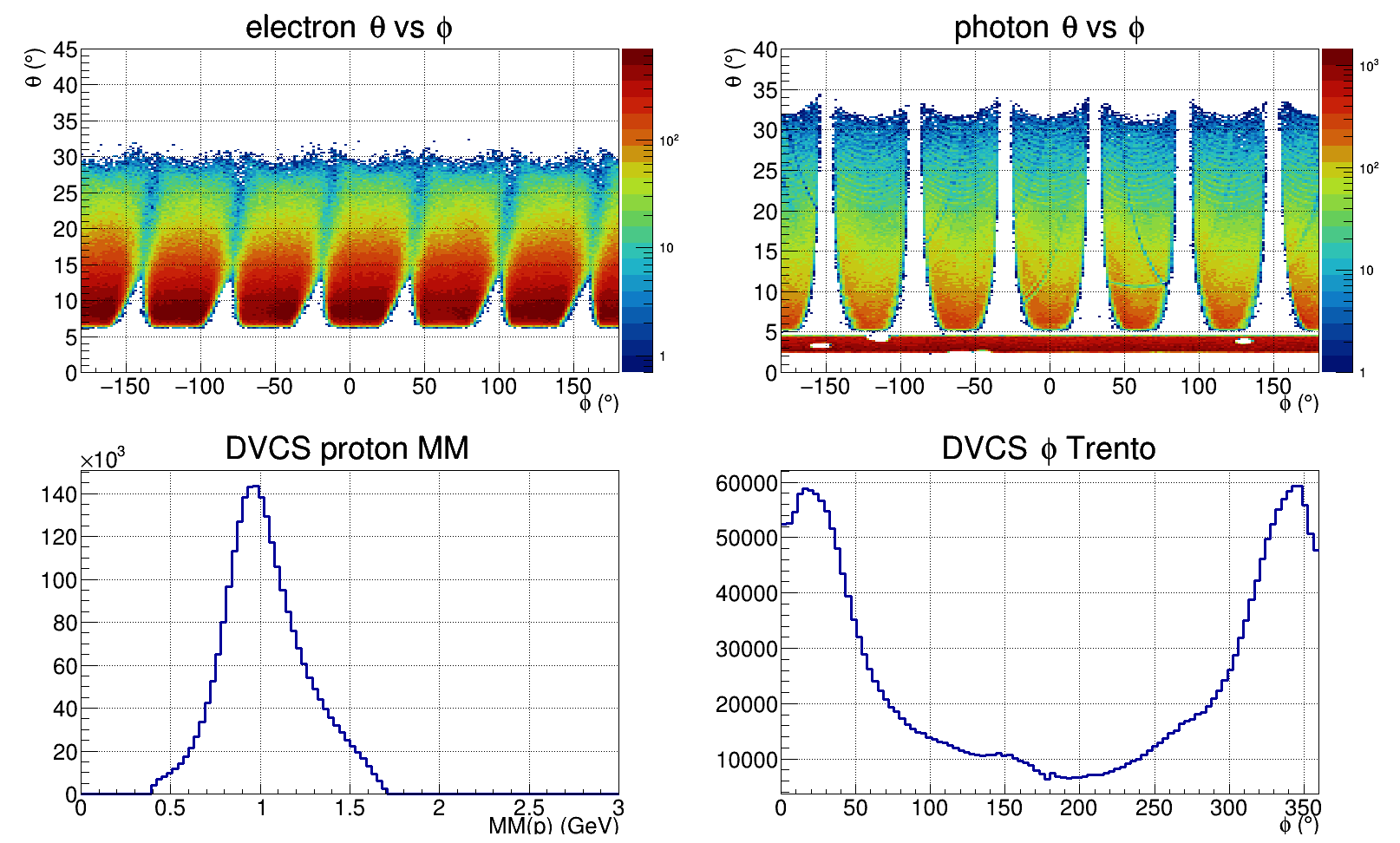}
\caption{DVCS/BH event distributions in individual Torus sectors: lepton $\phi$-distribution versus polar angle ($\theta)$ (top left), showing a slightly slanted asymmetric distribution due to the solenoid magnetic field; high energy photon distribution (top right), showing the ECAL (PCAL+EC) coverage at $\theta > 5^\circ$ and the FT calorimeter coverage at $\theta < 5^\circ$; the e$\gamma$X missing mass distribution, peaking at the proton mass (bottom left), with the radiative tail on the high mass side; the azimuthal distribution of DVCS events (bottom right).}
\label{analysis}
\end{center}
\end{figure}

The kinematics coverage is shown in Fig.~\ref{dvcs-kine}. Scattered leptons cover the $x_B$ range from 0.1 to 0.7 and a range in $Q^2$ from 1 to 10 GeV$^2$. The range in $-t$ covers 0.05 to 2.5~GeV$^2$. In Fig.~\ref{analysis} we show the event distribution in the individual FD sectors, the e$\gamma$X missing mass distribution with all particles detected, and the $\phi$-distributions of exclusive events.

\subsection{Monte Carlo simulations of background}

\begin{figure}[!t]
\centering
\includegraphics[width=.9\textwidth]{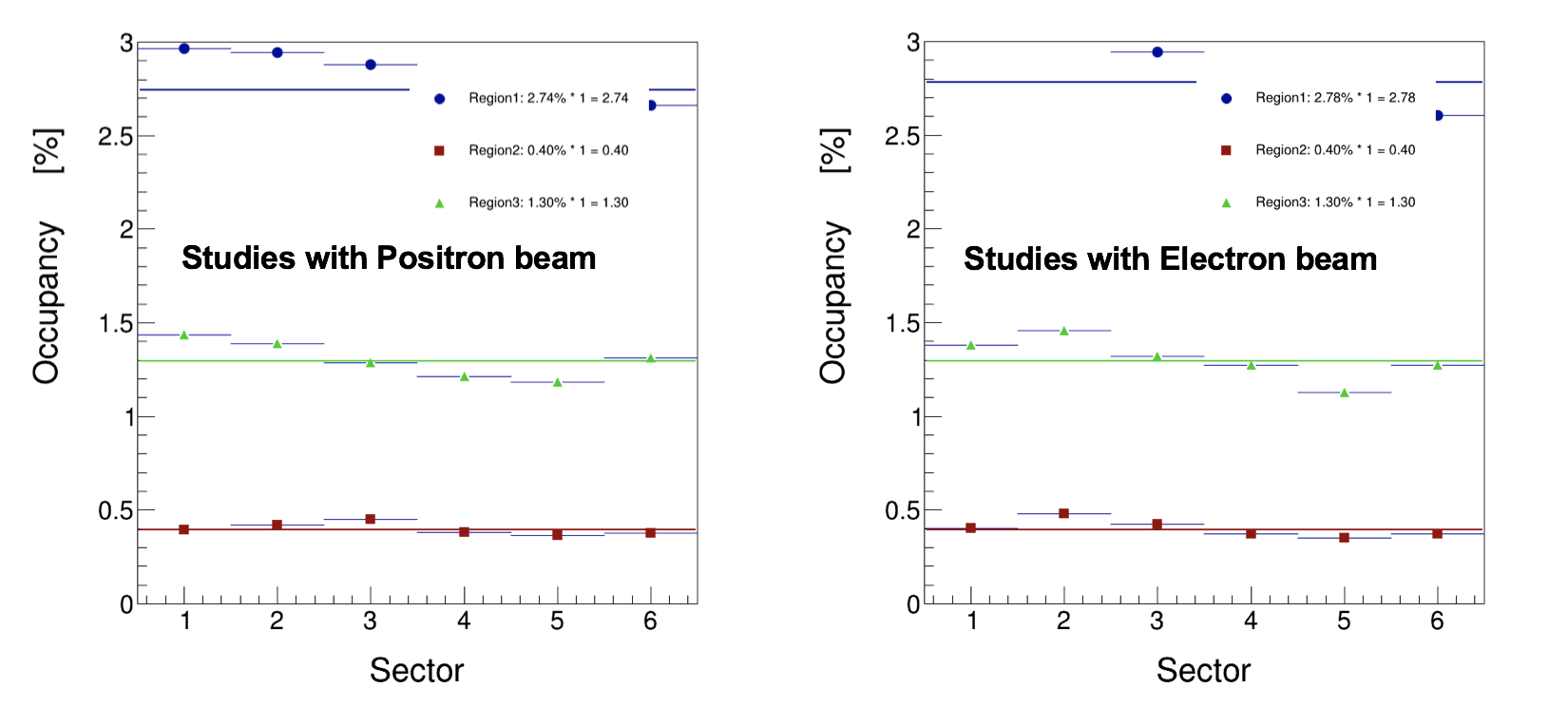}
\caption{Results for {\tt CLAS12} drift chambers occupancy obtained from GEANT simulations with positron and electron beams at a luminosity of 10$^{35}$cm$^{-2}$s$^{-1}$.}
\label{DCOccupancy}
\end{figure}

A critical part of operating {\tt CLAS12} detector at high luminosities is the simulation not only of hadronic events but also the simulation of beam-related accidental hits in the {\tt CLAS12} detector systems, in particular the tracking devices. Source of accidentals in this experiment is primarily from the positron elastically scattering off atomic electrons (Bhabha scattering) and their secondary interaction with beamline components.  The production rate of this background sources is orders of magnitude larger than the hadronic production rate. In the case of {\tt CLAS12} experiments with electron beam, the source of accidentals is primarily from beam electrons undergoing M{\o}ller scattering off atomic electrons in the liquid hydrogen target. The shielding of the {\tt CLAS12} detector from this background was carefully and extensively studied  during the {\tt CLAS12} design and construction. The final solution of the shielding was obtained by combination of magnetic shielding from the {\tt CLAS12} 5~T superconducting solenoid and carefully optimized design and fabrication of the M{\o}ller absorber. In studying the background for this experiment, detailed GEANT simulation were performed based on the {\tt CLAS12} realistic simulation package used for electron beam at luminosity of 10$^{35}$cm$^{-2}\cdot$s$^{-1}$ reversing the Torus field with respect to the electron beam configuration. We performed detailed comparison of the drifts chambers occupancy with results obtained with electron beam at the same luminosity taking into account the correct DC time windows of each region. The summary results of the simulation of the drift chamber occupancies are shown in figure~\ref{DCOccupancy}. Region 1 has the highest occupancy of about 3\%, region 2 is about 0.8\%, and region 3 is about 1.2\%. These results are compatible with the ones obtained with electron beam. 

In summary, no additional shielding is needed for this experiment, and we can switch between running the experiment with positrons and electrons keeping the {\tt CLAS12} configuration and the operating luminosity, the same. Additionally, a realistic simulation package is essential to take into account the detector occupancies for data taking at a given luminosity. In order to quantitatively account for this, pre-scaled trigger bits will be setup to take randomly triggered data simultaneously with the production data. Then in the offline data analysis we will merge DVCS simulation events with random triggered events from data to evaluate tracking and particle identification efficiency and monitor eventual differences between electron and positron running.

%
%

%
%
%
%
%

\section{Control of systematic uncertainties}
\label{sec:sys}

\subsection{Origin of systematic effects}

Systematic effects can occur on the one hand from positron beam properties, and on the other hand from the response of the {\tt CLAS12} spectrometer. 

The positron beam properties in the physics interaction region differ from the CEBAF electron beam essentially by a 4-5 times larger emittance~\cite{Rob17}. Additional focusing of the beam will be provided by a set of quadrupoles already installed on the beam line but residual effects of the beam properties difference may remain and alter the comparison with electron data. To control these effects DVCS data with an electron beam having the same properties as the positron beam should be carried out. Such beam can be made out of the secondary electrons produced at the positron source, which has similar properties in terms of $(x,y)$ profile and emittance. This will allow for the  elimination/correction of potential beam-related false asymmetries.

Systematic uncertainties for DVCS cross section measurements with electron beams are part of the currently ongoing program with the {\tt CLAS12} spectrometer setup. For positron beams most of these systematic effects are very similar as far as the {\tt CLAS12} detector properties and normalization issues are concerned. As we aim for measurements of charge asymmetries most of them will be identical and will therefore drop out in the ratio. However, there are a few effects that may not drop out and have to be considered.

\begin{figure}[!t]
\begin{center}
\includegraphics[width=0.50\columnwidth]{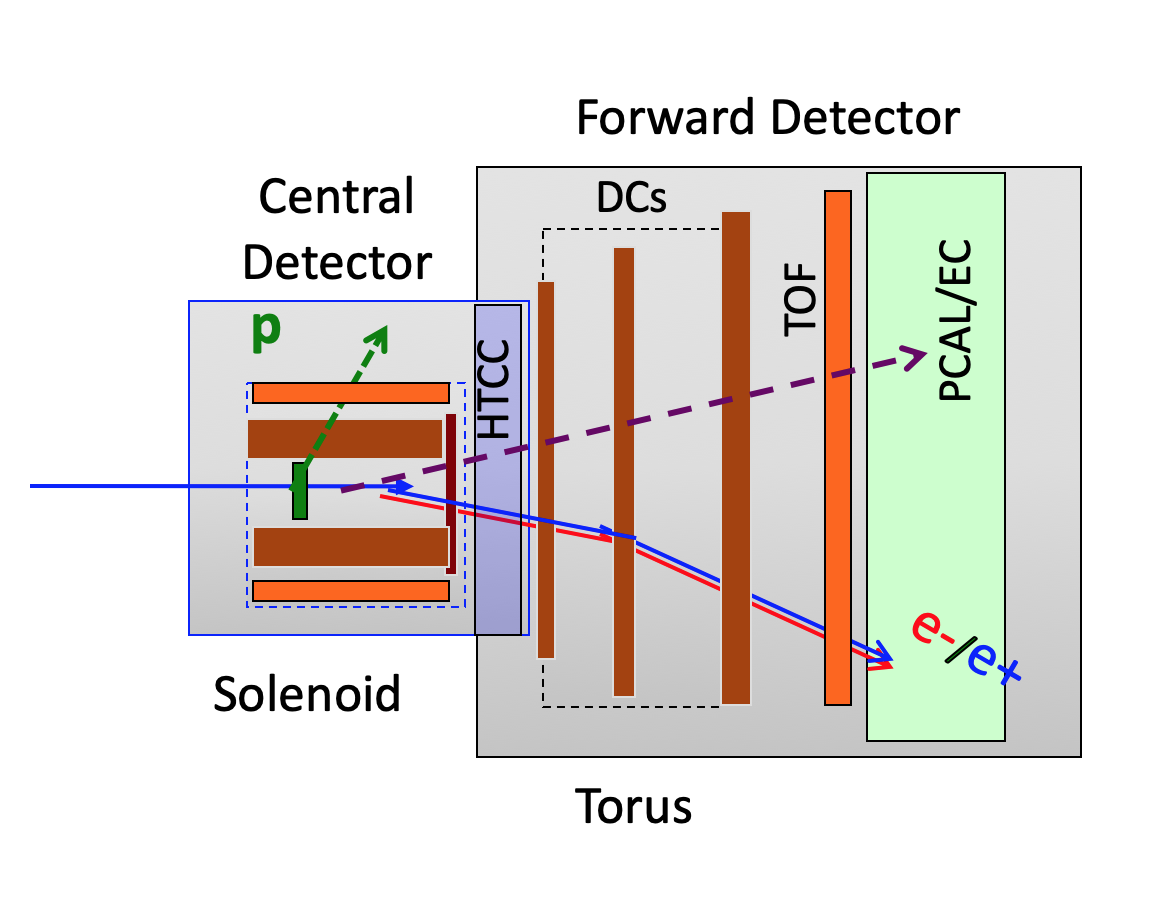}
\caption{The generic setup of the {\tt CLAS12} detector in Hall B in a side view. For the same kinematics the forward going particles electrons and positrons are following the same path for reversed polarity of the Torus and the solenoid magnets (both bending away from the beam line).} 
\label{CLAS12-side}
\end{center}
\end{figure} 
\begin{figure}[!t]
\begin{center}
\includegraphics[width=0.70\columnwidth]{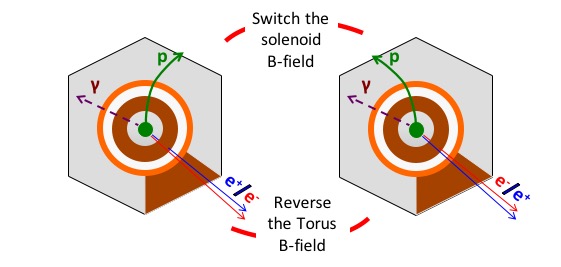}
\caption{The generic setup of the {\tt CLAS12} detector in Hall B viewed from upstream down the beam pipe. In this view the proton rotates in the opposite direction, from the case of the electron beam. When switching the Torus field electrons and positrons experience different phi motions due to the solenoid (left). When the solenoid field is reversed electrons and positrons get kicks in the opposite azimuthal directions and positrons and electrons switch place in the FD, but protons are detected in another CD region. This may lead to false asymmetries that can be controlled with proton elastic scattering.} 
\label{clas12-cartoon}
\end{center}
\end{figure}

The experimental setup is generically shown in a side view in Fig.~\ref{CLAS12-side}, and in a view along the beam line looking downstream in Fig.~\ref{clas12-cartoon}. For scattered positrons and for the DVCS photons the detector looks identical to the situation when electrons are scattered off protons and the magnetic fields in both magnets are reversed. This is not the case of recoil protons, which will be bent in the solenoid field in the opposite direction compared to the electron scattering case. This could result in systematic effects due to potentially different track reconstruction efficiencies and effective solid angles. While these effects are much smaller than the physics asymmetry, we propose to measure the DVCS process 50\% of the time with reversed solenoid polarity. Simultaneously to DVCS, the elastic $e^{\pm}p \to e^{\pm}p$ scattering cross section will be also measured continuously during the experiments. This will provide additional cross check and a monitoring tool of the performance of the detector system throughout the experiment. It should be noted that the elastic $e^+p \to e^+p$ cross section is equal to the well known elastic $e^-p \to e^-p$ cross section within the contributions from 2-photon effects, which are very small in the kinematical range that can be selected for the process~\cite{Afa17}.  

From the simulation calibrated by the measurement of known processes, we aim at keeping systematic uncertainties due to detector asymmetries below 5\% at all kinematics. We also note that for the positron BSA, which will be measured simultaneously, the systematic uncertainties are not affected by the aforementioned electron-positron differences. 

\subsection{Beam charge asymmetry systematics}

\begin{figure}[!t]
\begin{center}
\includegraphics[width=0.85\textwidth]{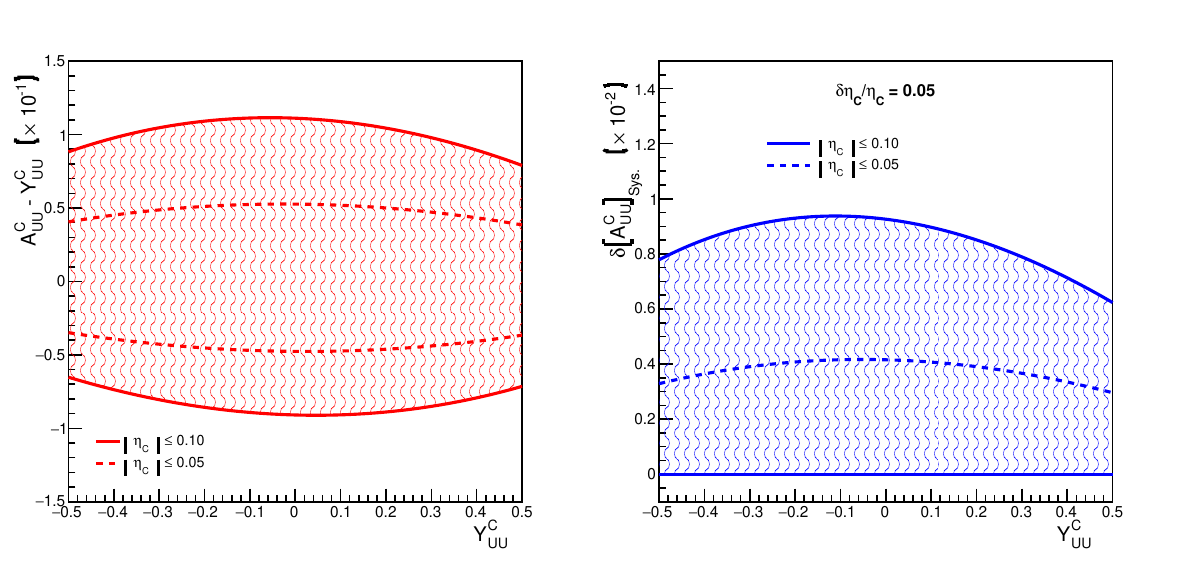}
\includegraphics[width=0.85\textwidth]{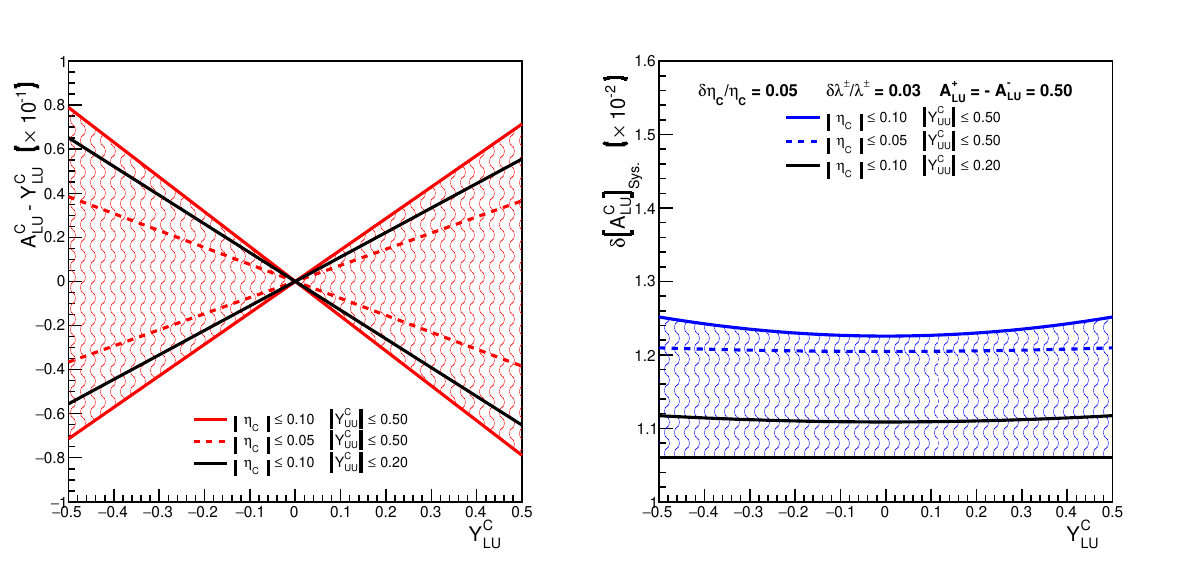}
\includegraphics[width=0.85\textwidth]{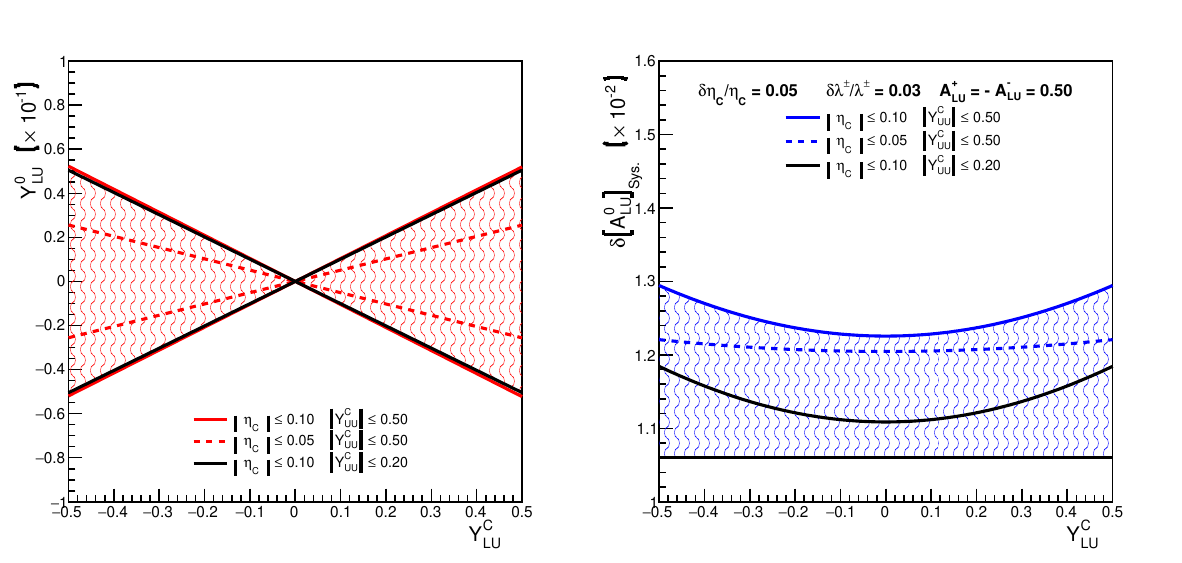}
\caption{Amplitude of the corrections of raw asymmetries (left) and of the corresponding systematic errors (right). The shaded areas indicate the possible values of the corrections for a fixed raw asymmetry assuming that detector difference effects are comprised within $\pm$10\% (solid line) or $\pm$5\% (dashed line). Asymmetry systematics is evaluated assuming 5\% relative uncertainty on the detector correction factor $\eta_C$ and twist-2 approximation for polarized BCAs. The solid black line limits the variation domain when the unpolarized BCA is comprised within $\pm$20\%.}
\label{Sys_BCA}
\end{center}
\end{figure}

The unpolarized raw Yield Charge Asymmetry (YCA) is defined experimentally as
\begin{equation}
\mathcal{Y}^C_{UU} = \frac{(Y^+_+ + Y^+_-) - (Y^-_+ + Y^-_-)}{Y^+_+ + Y^+_- + Y^-_+ + Y^-_-}
\end{equation}
where 
\begin{equation}
Y^e_{\lambda} = \frac{N^e_{\lambda}}{Q^e_{\lambda}} = n^e_{\lambda}
\end{equation}
is the beam charge and spin dependent normalized yield determined from the number of events $N^e_{\lambda}$ and the corresponding accumulated charge $Q^e_{\lambda}$. Comparing electron and positron measurements taken at different periods of time, the detector may not be exactly the same. This translates into different efficiencies ($\epsilon^{\pm}$) and solid angles ($\Delta \Omega^{\pm}$) leading to an excess/deficit of positron events as compared to the true physics event number expected within the acceptance of electron beam data. Noting $\delta n^0_0$ the spin and charge independent positron excess, the charge normalized yield for electrons and positrons can be expressed as
\begin{eqnarray}
n^-_{\pm} & = & n^0_0 \left[ 1 - A^C_{UU} \pm \lambda^- \left( A^0_{LU} - A^C_{LU}\right) \right] \\
n^+_{\pm} & = & n^0_0 \left( 1 + \frac{\delta n^0_0}{n^0_0} \right) \, \left[ 1 + A^C_{UU} \pm \lambda^+ \left( A^0_{LU} + A^C_{LU}\right) \right] 
\end{eqnarray}
where $n_0^0$ is the unpolarized normalized neutral yield for identical detectors and the $A^i_{jk}$ are the physics BCAs. The detector difference is represented by the ratio
\begin{equation}
\frac{\epsilon^+ \Delta \Omega^+}{\epsilon^- \Delta \Omega^-} = 1 + \frac{\delta n_0^0}{n^0_0} = 1 + 2 \eta_C
\end{equation}
where $\eta_C$ represents the detector differences correction. The unpolarized physics BCA can be derived from the experimental YCA following the expression 
\begin{equation}
A^C_{UU} = \frac{\left( 1 + \eta_C \right) \, \mathcal{Y}^C_{UU} - \eta_C}{1 + \eta_C - \eta_C \, \mathcal{Y}^C_{UU}} \, . \label{ACuu:YCUU}
\end{equation}
Similarly, the polarized raw YCAs are defined as
\begin{eqnarray}
\mathcal{Y}^C_{LU} & = & \frac{(Y^+_+ - Y^+_-)/\lambda^+ - (Y^-_+ - Y^-_-)/\lambda^-}{Y^+_+ + Y^+_- + Y^-_+ + Y^-_-} \\
\mathcal{Y}^0_{LU} & = & \frac{(Y^+_+ - Y^+_-)/\lambda^+ + (Y^-_+ - Y^-_-)/\lambda^-}{Y^+_+ + Y^+_- + Y^-_+ + Y^-_-}
\end{eqnarray}
from which the polarized physics BCAs are obtained as  
\begin{eqnarray}
A^C_{LU} & = & \frac{(1+\eta_C) \mathcal{Y}^C_{LU} - \eta_C \mathcal{Y}^0_{LU}}{1+\eta_C-\eta_C \, \mathcal{Y}^C_{UU}} \approx \frac{1+2\eta_C}{1+\eta_C} \, \frac{\mathcal{Y}^C_{LU}}{1+\eta_C-\eta_C \, \mathcal{Y}^C_{UU}} \label{ACLu:YCLU} \\
A^0_{LU} & = & \frac{(1+\eta_C) \mathcal{Y}^0_{LU} - \eta_C \mathcal{Y}^C_{LU}}{1+\eta_C-\eta_C \, \mathcal{Y}^C_{UU}} \approx 0 \label{A0Lu:Y0LU}
\end{eqnarray}
where the approximation holds at twist-2. Because of possible detector asymmetry effects between beams of opposite charges, polarized BCAs combine unpolarized and polarized raw asymmetries. Thanks to the polarity reversal of the {\tt CLAS12} magnets, the magnitude of these effects will be minimized and it is reasonable to expect differences smaller than 10\%. The magnitude of the corrections to raw unpolarized and polarized YCAs are determined as the deviation from the physics BCAs. The left panel of Fig.~\ref{Sys_BCA} reports the envelope of the possible corrections assuming $\eta_C$ values within $\pm$0.05 (or $\pm$0.02) and the twist-2 approximation for polarized BCAs. Detector effects might be sizeable but are easily corrected once their value is determined. 

The precision on these corrections directly enters the systematic errors on BCAs, which are determined as
\begin{eqnarray}
\delta \left[ A^C_{UU} \right]_{Sys.} & = & \frac{\left\vert  { 1 - {\left[ \mathcal{Y}^C_{UU} \right]}^2 } \right\vert}{\left( {1 + \eta_C - \eta_C \, \mathcal{Y}^C_{UU}} \right)^2} \, \delta \eta_C \\
\delta \left[ A^C_{LU} \right]_{Sys.} & = & {\left[  {\left( \frac{\mathcal{Y}^C_{LU} \mathcal{Y}^C_{UU} - \mathcal{Y}^0_{LU}}{{\left( 1 + \eta_C - \eta_C \, \mathcal{Y}^C_{UU} \right)}^2} \, \delta \eta_C \right)}^2 +
\sum_{i=\pm}{\left( \frac{\partial A^C_{LU}}{\partial \lambda^i} \, \delta\lambda^i \right)}^2 \right]}^{1/2}\label{eq:ACLU_err} \\
\delta \left[ A^0_{LU} \right]_{Sys.} & = & {\left[  {\left( \frac{\mathcal{Y}^0_{LU} \mathcal{Y}^C_{UU} - \mathcal{Y}^C_{LU}}{ {\left( 1 + \eta_C - \eta_C \, \mathcal{Y}^C_{UU} \right)}^2} \, \delta\eta_C \right)}^2 + \sum_{i=\pm}{\left( \frac{\partial A^0_{LU}}{\partial \lambda^i} \, \delta\lambda^i \right)}^2 \right]}^{1/2} \label{eq:A0LU_err}
\end{eqnarray} 
where the additional contributions in Eq.~\ref{eq:ACLU_err}-\ref{eq:A0LU_err} take into account the beam polarization systematics according to the relations
\begin{eqnarray}
& & \frac{\partial A^C_{LU}}{\partial \lambda^{+}} \, \delta\lambda^{+} = \frac{\partial A^0_{LU}}{\partial \lambda^{+}} \, \delta\lambda^{+} = \frac{1}{2} \, \frac{A^{+}_{LU}(1 + \mathcal{Y}^C_{UU})}{1 + \eta_C - \eta_C \, \mathcal{Y}^C_{UU}} \, \frac{\delta\lambda^{+}}{\lambda^{+}} \\
& & \frac{\partial A^C_{LU}}{\partial \lambda^{-}} \, \delta\lambda^{-} = - \frac{\partial A^0_{LU}}{\partial \lambda^{-}} \, \delta\lambda^{-} =  
\frac{1+2\eta_C}{2} \, \frac{A^{-}_{LU}(1 - \mathcal{Y}^C_{UU})}{1 + \eta_C - \eta_C \, \mathcal{Y}^C_{UU}} \, \frac{\delta\lambda^{-}}{\lambda^{-}} \, .
\end{eqnarray}

\newpage

\null\vfill

\vfill\eject

The right panel of Fig.~\ref{Sys_BCA} reports the envelope of systematic errors related to detector difference effects for unpolarized and polarized BCAs, assuming $\delta \eta_C / \eta_C$=0.05 and $\delta \lambda^+ / \lambda^+$=$\delta \lambda^- / \lambda^-$=0.03. Furthermore, the polarized BCAs systematic are determined using the twist-2 approximation and maximizing the contribution of the beam polarization systematic choosing $A^+_{LU}$=-$A^-_{LU}$=0.50. Different domains of $\eta_C$ and $\mathcal{Y}^C_{UU}$ values are considered 
which define the correction systematic envelopes. The absolute systematics of unpolarized BCAs is smaller than 0.010, and 0.013 for polarized BCAs. These values can be considered as the minimal BCAs that can be meaningfully measured.

%
%

%
%
%
%
%

\section{Beam time request}

For the main physics program we request 80 days of 5-pass secondary electron and positron beams at an energy of 10.6 GeV, and a beam current of 50~nA impinging on a 5~cm long liquid hydrogen target. The beam must be longitudinally polarized with a polarization of $\lambda$=$60$\% or higher. The beam polarization must be switched at a rate of 30~Hz with direct reporting, and the charge asymmetry between each helicity state must be kept below 0.1\%. These figures are supplemented with 8 days of background calibration studies. In addition, we request 6 days of 1-pass electron beam at the same current of 50~nA for systematic effects and cross calibration purposes of the beam line instrumentation, and 6 days of commissionning of the Hall B equipment with the positron beam.

\subsection{Proposed measurements}

The measurements of this proposal are detailed in Tab.~\ref{Btr} and comprise physics data taking with both electron and positron beams. Electron data using the secondary electron beam generated at the positron production target will be used to compare with positron data for constructing BCA observables. Comparing electron and positron data taken with similar statistics, the same beam properties, the same target cell, the same detector status, and the same detector configuration is the path we propose to miminize systematic effects and provide high quality BCA observables.

Immediately before the beginning of the physics run, short electron beam runs will be performed to enable checking for false charge asymmetries by employing elastic $e^-p \to e^-p$ scattering with a 1-pass beam, and commission the polarimeter in electron mode. This comprises using both the primary Ce$^+$BAF electron beam and the secondary beam generated at the positron production target, together with negative and positive solenoid polarities. The electron beam physics run will be conducted over a period of 40 days, alternating solenoid polarities and using the secondary electron beam produced simultaneously with the positron beam at the production target of the positron source. \newline
After the polarity change of CEBAF, the Hall B beamline equipment and polarimeter will be commissioned with the positron beam at 1-pass, and a short run on the $e^+p \to e^+p$ elastic scattering will be performed for direct comparison with electron calibration data. For this comparison elastic kinematics at low $Q^2$ will be selected to limit the 2-photon contributions to less than 2-3\%. After commissioning Hall B equipment with a 5-pass positron beam, the $e^+p \to e^+p\gamma$ DVCS physics run will be conducted. \newline
Polarization measurements will be carried out regularly every 2-3 days initially, and once a week during stable running periods. The current plan is to use Bhabha scattering on a polarized ferromagnetic foil (see Appendix~\ref{App-A}). \newline
Additional calibration runs will also be conducted all along the physics run for detector calibration and background studies. Production data taking is expected with a trigger rate of up to 20~KHz and a data rate of up to 800~MB/s. Once a week during stable operation, luminosity scan will be carried out and randomly triggered data will be taken at various beam currents to simulate realistic background conditions to be used for Monte Carlo simulations. \newline
The total beam time request amounts to 2400 hours (100 days) of data taking. The present data taking scenario would benefit for more frequent switch between electron and positron beams in order to minimize the effects of an eventual drift of the detector response. A fast (a few shifts) switch of the polarity of the CEBAF magnets would be mandatory for this purpose. 

\begin{table}[!t]
\begin{center}
\resizebox{0.83\textwidth}{!}{
\begin{tabular}{@{}l|c||c|c|c|c|c||c|c|c||r@{}} \hline \hline

\multirow{3}{*}{Purpose} & \multirow{3}{*}{Label} & \multicolumn{5}{c||}{Beam parameters} & \multirow{3}{*}{Target} & \multirow{2}{*}{Sol.} & \multirow{2}{*}{Tor.} & \multirow{3}{*}{Time} \\
 & & q & \multirow{2}{*}{Nat.} & E & I & $\lambda$ & & \multirow{2}{*}{Pol.} & \multirow{2}{*}{Pol.} & \\
 & & (e) & & (GeV) & (nA) & (\%) & & & & (h)\phantom{d} \\ \hline \hline

\multirow{2}{*}{$ep \to ep$} & \multirow{2}{*}{Cal.} & \multirow{9}{*}{$-$} & \multirow{2}{*}{P} & \multirow{5}{*}{2.2} & \multirow{9}{*}{50} & \multirow{5}{*}{0} &  & $-$ & \multirow{9}{*}{$+$} & 24 \\ \cline{9-9}\cline{11-11}
& & & & & & & & $+$ & & 24 \\ \cline{1-2} \cline{4-4} \cline{9-9} \cline{11-11}

 \multicolumn{2}{l||}{Commissioning} & & \multirow{7}{*}{S} & & & & & $+$ & & 24 \\ \cline{1-2} \cline{9-9} \cline{11-11}
  
\multirow{2}{*}{$ep \to ep$} & \multirow{2}{*}{Cal.} & & & & & & & $+$ & & 24 \\ \cline{9-9}\cline{11-11}

 & & & & & & & 5~cm & $-$ & & 24 \\ \cline{1-2} \cline{5-5} \cline{7-7} \cline{9-9} \cline{11-11}

$ep \to ep\gamma$ & Phy. & & & \multirow{4}{*}{10.6} & & \multirow{4}{*}{60} & LH$_2$ & $-$ & & 480 \\ \cline{1-2} \cline{9-9} \cline{11-11}
Background &  Cal.& & & & & & & $-$ & & 48 \\ \cline{1-2} \cline{9-9} \cline{11-11}
$ep \to ep\gamma$ & Phy. & & & & & & & $+$ & & 480 \\ \cline{1-2} \cline{9-9} \cline{11-11}
Background & Cal. & & & & & & & $+$ & & 48 \\ \cline{1-3} \cline{5-5} \cline{9-11} \hline\hline
\multicolumn{2}{l||}{Commissioning} & \multirow{8}{*}{$+$} & \multirow{8}{*}{S} & \multirow{3}{*}{2.2} & \multirow{8}{*}{50} &  \multirow{3}{*}{0} & & $+$ & \multirow{8}{*}{$-$} & 48 \\ \cline{1-2} \cline{9-9} \cline{11-11}
\multirow{2}{*}{$ep \to ep$} & \multirow{2}{*}{Cal.} & & & & & & & $+$ & & 24 \\ \cline{9-9} \cline{11-11}
& & & & & & & & $-$ & & 24 \\ \cline{1-2} \cline{5-5} \cline{7-7} \cline{9-9} \cline{11-11}
\multicolumn{2}{l||}{Commissioning} & & & \multirow{5}{*}{10.6} & & \multirow{4}{*}{60} & 5~cm & $-$ & & 72 \\ \cline{1-2} \cline{9-9} \cline{11-11}
 $ep \to ep\gamma$ & Phy. & & & & & & LH$_2$ & $-$ & & 480 \\ \cline{1-2} \cline{9-9} \cline{11-11}
Background & Cal. & & & & & & & $-$ & & 48 \\ \cline{1-2} \cline{9-9} \cline{11-11}
 $ep \to ep\gamma$ & Phy. & & & & & & & $+$ & & 480 \\ \cline{1-2} \cline{9-9} \cline{11-11}
Background & Cal. & & & & & & & $+$ & & 48 \\ \hline\hline
\multicolumn{10}{r||}{\bf Total} & {\bf 2400} \\ 
\end{tabular}
}
\caption{Detailed description of the beam time request. The beam nature label P indicates the CEBAF primary electron beam, and S indicates the secondary electron or positron beam generated at the positron production target. The data taking label Cal. stands for calibration, and Phy. for physics data taking.}
\label{Btr}
\end{center}
\end{table}

\subsection{Experimental projections}

\begin{figure}[!h]
\begin{center}
\includegraphics[width=0.9\textwidth]{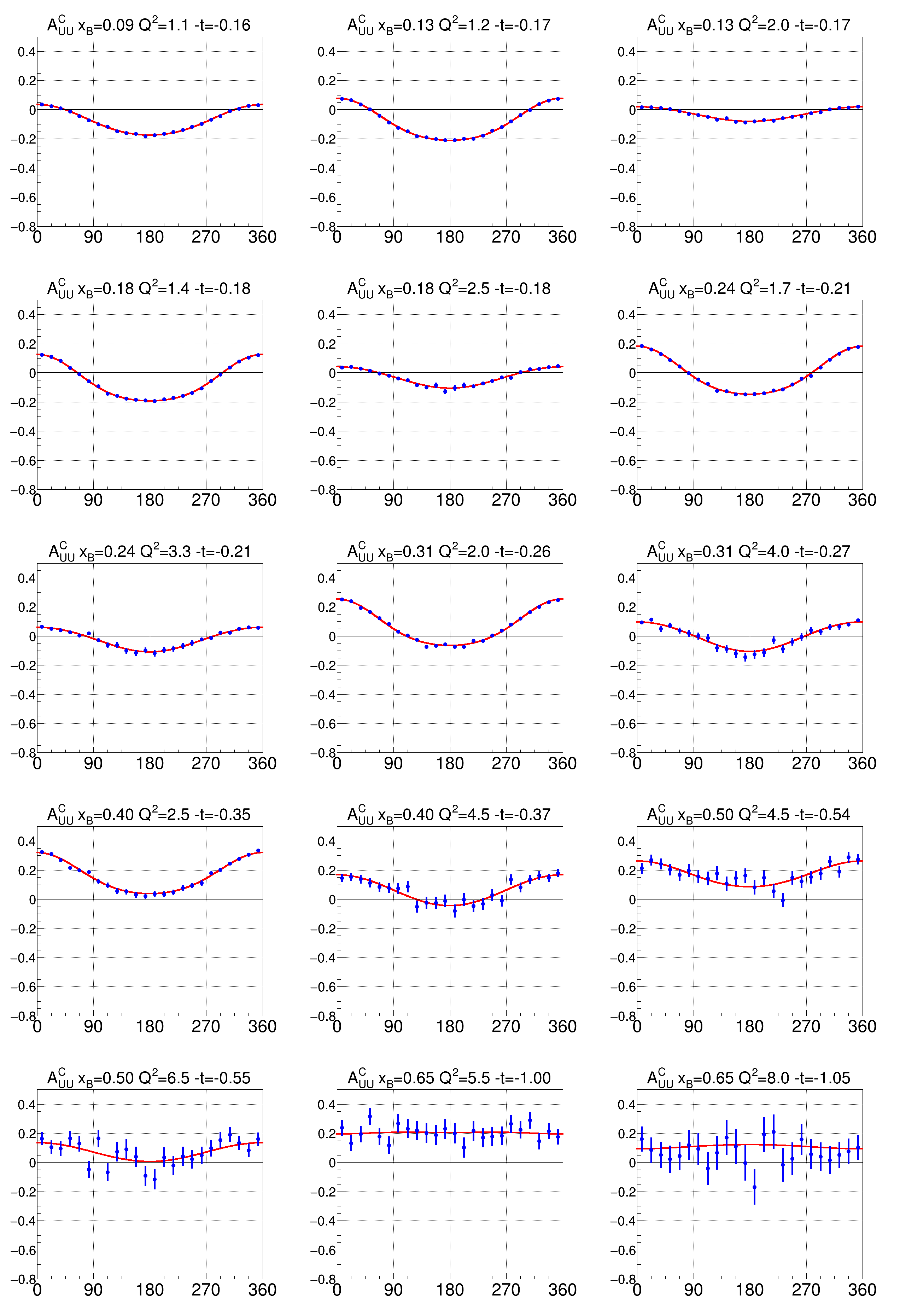}
\caption{Projected $A^C_{UU}$ data in selected small $t$-bins at a 10.6 GeV positron and electron beam energy.}
\label{BCA_t2}
\end{center}
\end{figure}

\begin{figure}[!h]
\begin{center}
\includegraphics[width=0.9\textwidth]{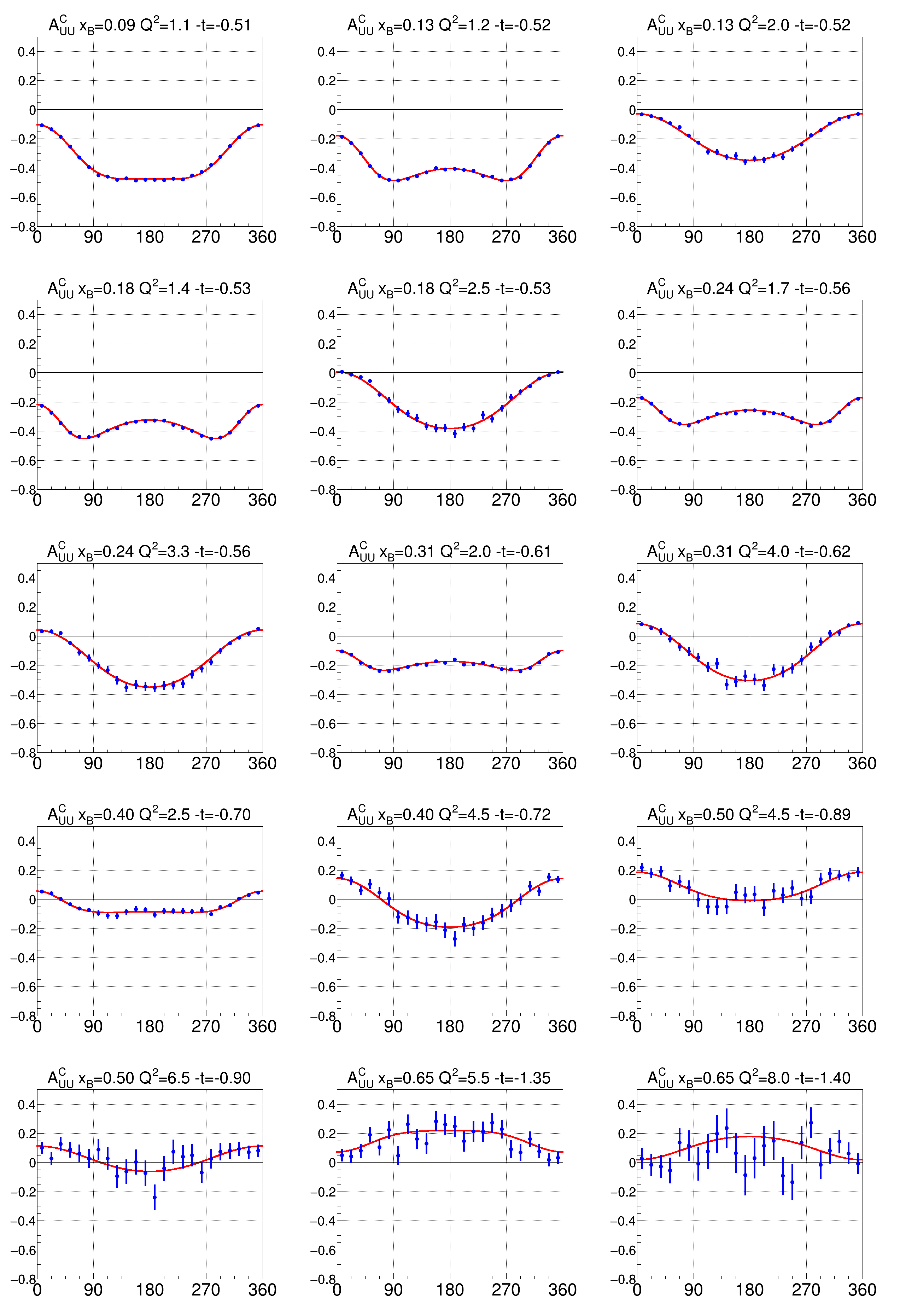}
\caption{Projected $A^C_{UU}$ data in selected intermediate $t$-bins at a 10.6 GeV positron and electron beam energy.}
\label{BCA_t4}
\end{center}
\end{figure}

\begin{figure}[!h]
\begin{center}
\includegraphics[width=0.9\textwidth]{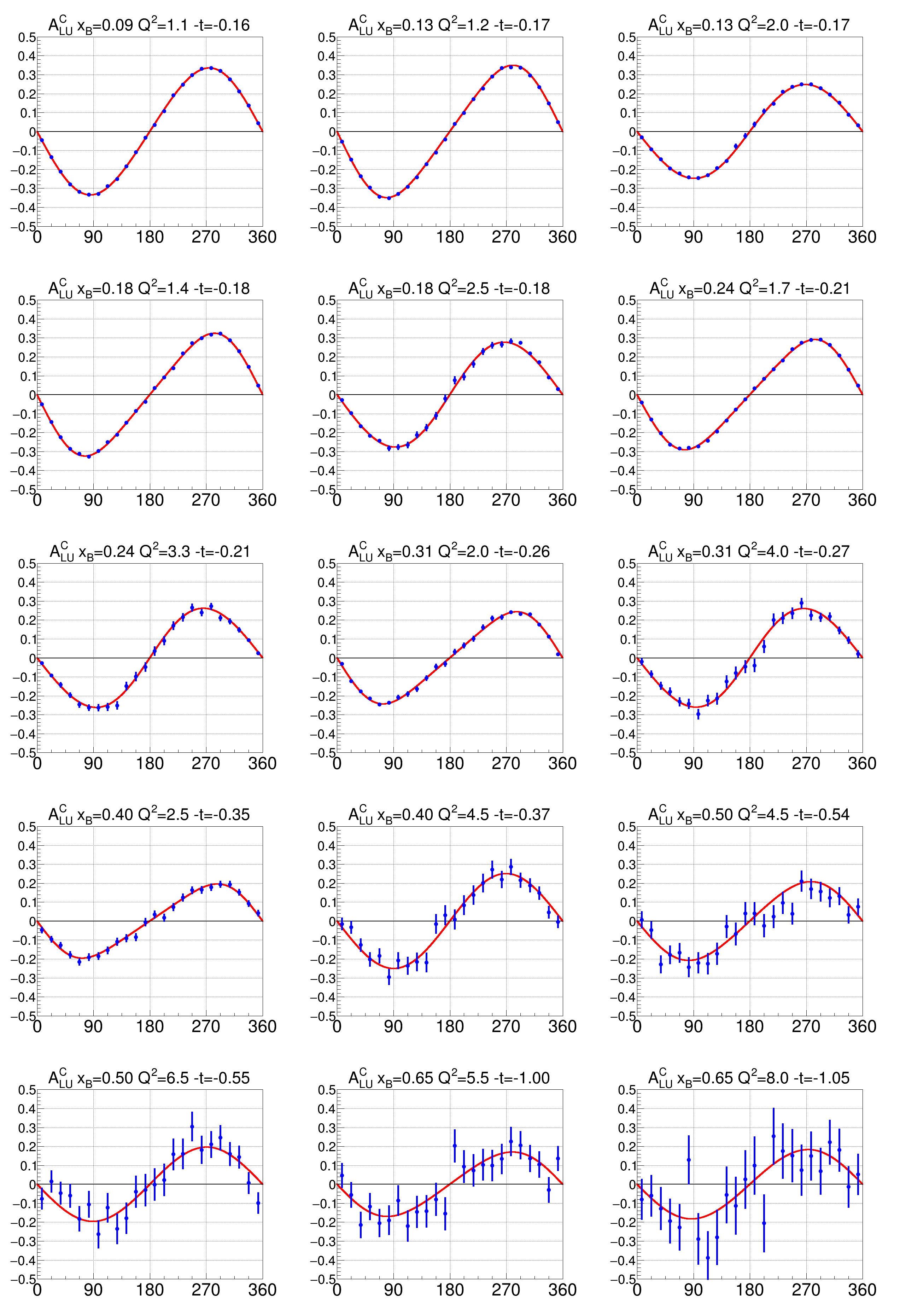}
\caption{Projected $A^C_{LU}$ data in selected small $t$-bins at a 10.6 GeV positron and electron beam energy.}
\label{BCSA_t2}
\end{center}
\end{figure}

\begin{figure}[!h]
\begin{center}
\includegraphics[width=0.9\textwidth]{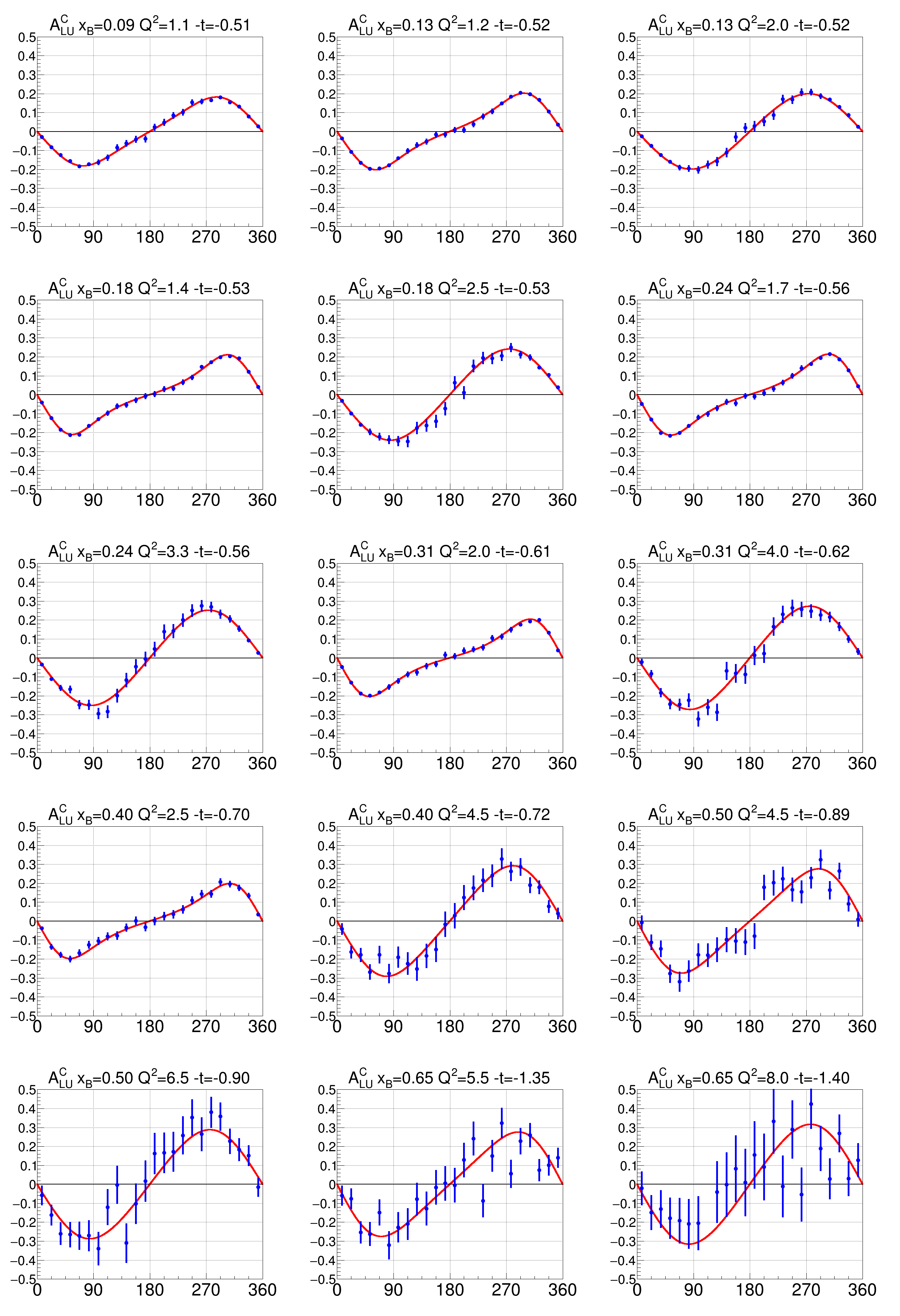}
\caption{Projected $A^C_{LU}$ data in selected intermediate $t$-bins at a 10.6 GeV positron and electron beam energy.}
\label{BCSA_t4}
\end{center}
\end{figure}

\begin{figure}[!h]
\begin{center}
\includegraphics[width=0.9\textwidth]{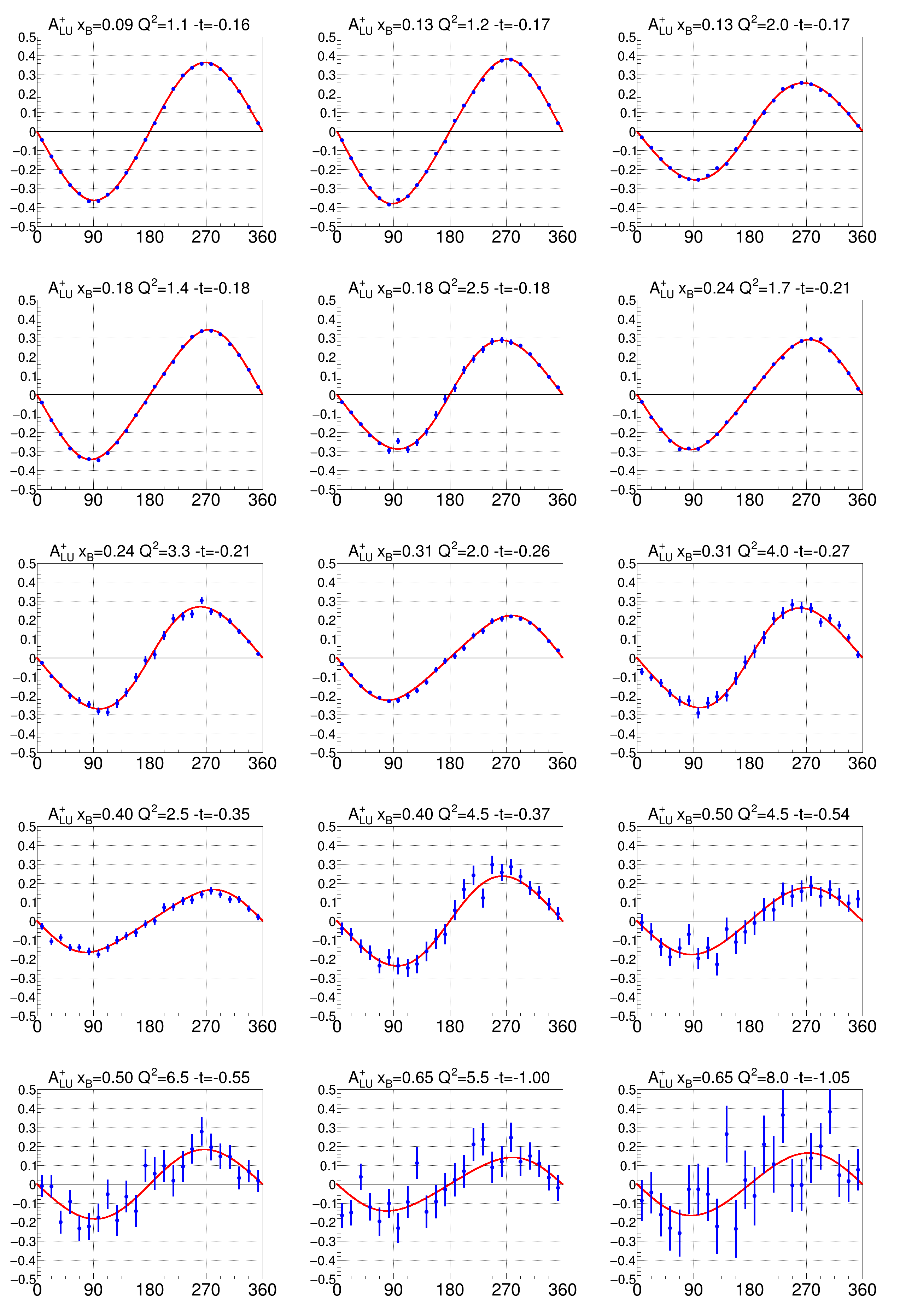}
\caption{Projected $A^+_{LU}$ data in selected small $t$-bins at a 10.6 GeV positron beam energy.}
\label{pBSA_t2}
\end{center}
\end{figure}

\begin{figure}[!h]
\begin{center}
\includegraphics[width=0.9\textwidth]{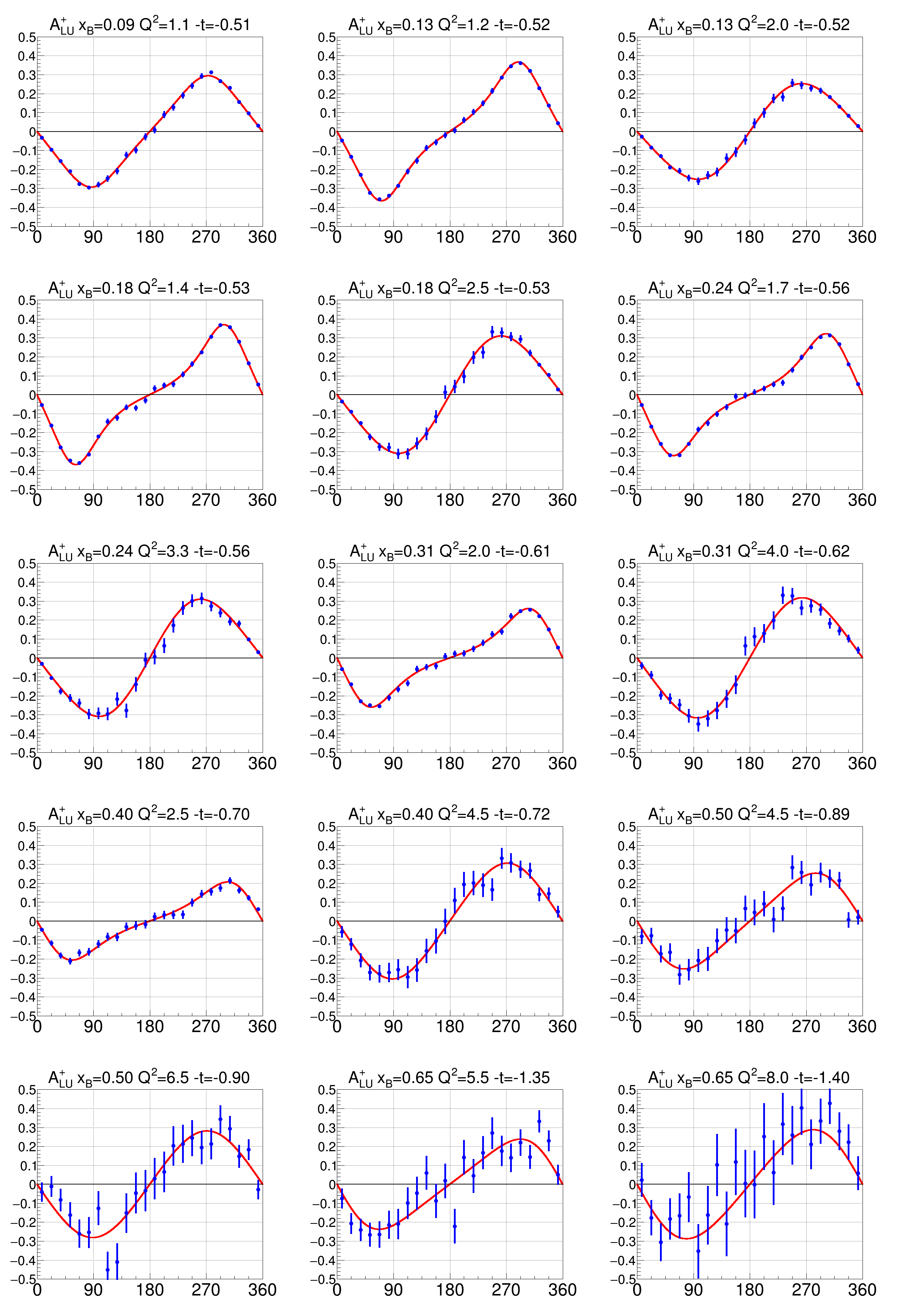}
\caption{Projected $A^+_{LU}$ data in selected intermediate $t$-bins at a 10.6 GeV positron beam energy.}
\label{pBSA_t4}
\end{center}
\end{figure}

Expected experimental data are shown in Fig.~\ref{BCA_t2}-\ref{pBSA_t4} at small and intermediate $t$ using the BM modeling of DVCS observables~\cite{Bel10} and the KM CFF~\cite{Kum10}, for the binning used in Sec.~\ref{Sec:Imp} to evaluate the impact of positron measurements on the CFF extraction. Statistical error bars assume 80 days DVCS data taking equaly shared between secondary electron and positron beams. The data represented in these figures correspond to a selected set of the data that will be obtained. Typical $\phi$-distributions of $A^C_{UU}$ (Fig.~\ref{BCA_t2}-\ref{BCA_t4}), $A^C_{LU}$ (Fig.~\ref{BCSA_t2}-\ref{BCSA_t4}), and $A^+_{LU}$  (Fig.~\ref{pBSA_t2}-\ref{pBSA_t4}) are shown for different $(x_B,Q^2)$-bins. The $(x_B,Q^2,t)$ sensitivity of the magnitude and shape of $A^C_{UU}$ is particularly noticed. In general, the accuracy of expected data will allow us to identify and quantitify the $\phi$-modulation of DVCS observables with a precision which decreases as $(x_B,Q^2)$ increase.

\newpage

\null\vfill

\vfill\eject

\null\vfill

\vfill\eject

\null\vfill

\vfill\eject

\null\vfill

\vfill\eject

\null\vfill

\vfill\eject

\null\vfill

\vfill\eject

%
%

%
%
\begin{appendix}

%
%
%
%
%

\newpage

%
%

\section{Beam polarization measurement}
\label{App-A}

The positron beam polarization can be measured using Bhabha scattering, where an incident positron scatters off an electron in a polarized metallic target. The most likely option for the design of a Bhabha polarimeter is to modify the existing M{\o}ller polarimeter to accommodate positrons. Though the cross section for Bhabha and M{\o}ller scattering are different, the analyzing power ($A_{zz}$) for a longitudinally polarized beam and a longitudinally polarized target are the same~\cite{Ale02}
\begin{equation}
\label{eq-Azz}
	A_{zz}\left(\theta_{CM}\right) = -\frac{\left(7+\cos\theta_{CM}\right)\sin^2\theta_{CM}}{\left(3+\cos^2\theta_{CM}\right)^2}
\end{equation}
where $\theta_{CM}$ is the center of momentum scattering angle. Particularly, $A_{zz}$ has a maximum magnitude of $7/9$ at $\theta_{CM}=90^\circ$, which is the central  scattering angle of the existing M{\o}ller polarimeter. Forming the beam-helicity-dependent asymmetry gives
\begin{equation}
\label{eq-asymm}
{\mathcal A} = \left( \frac{d\sigma^+}{d\Omega}-\frac{d\sigma_-}{d\Omega} \right)  \bigg{/} \left( \frac{d\sigma^+}{d\Omega}+\frac{d\sigma_-}{d\Omega} \right) = A_{zz} \left(\theta_{CM}\right) \, P_B^z P_T^z,
\end{equation}
where the $\pm$ refers to cases where the beam polarization ($P_B^z$) and the target polarization ($P_T^z$) are aligned or anti-aligned. The asymmetry is measured from the yields according to
\begin{equation}
\label{eq-asymm-meas}
{\mathcal A} = \frac{N_+-N_-}{N_++N_-} = \langle A_{zz}\rangle \, P_B^z P_T^z \, ,
\end{equation}
where $\langle A_{zz}\rangle$ is the effective analyzing power corrected for the finite-angle acceptance of the polarimeter and atomic-electron motion (also known as the Levchuk effect~\cite{Lev94}). The CLAS12 M{\o}ller polarimeter detects the scattered electrons in coincidence near $\theta_{CM}=90^\circ$, the peak of $A_{zz}$. As compared to single-arm M{\o}ller polarimetry, the coincidence method has the advantage of producing a clean data set without having to do energy-dependent background subtractions (see, for example Ref.~\cite{Arr92}). Accidental background rates are typically less than 10\% of the real coincident rate at the CLAS12 polarimeter, and is further measured and included as a correction.

\begin{figure}[!htb]
\begin{center}
\includegraphics[width=0.725\textwidth]{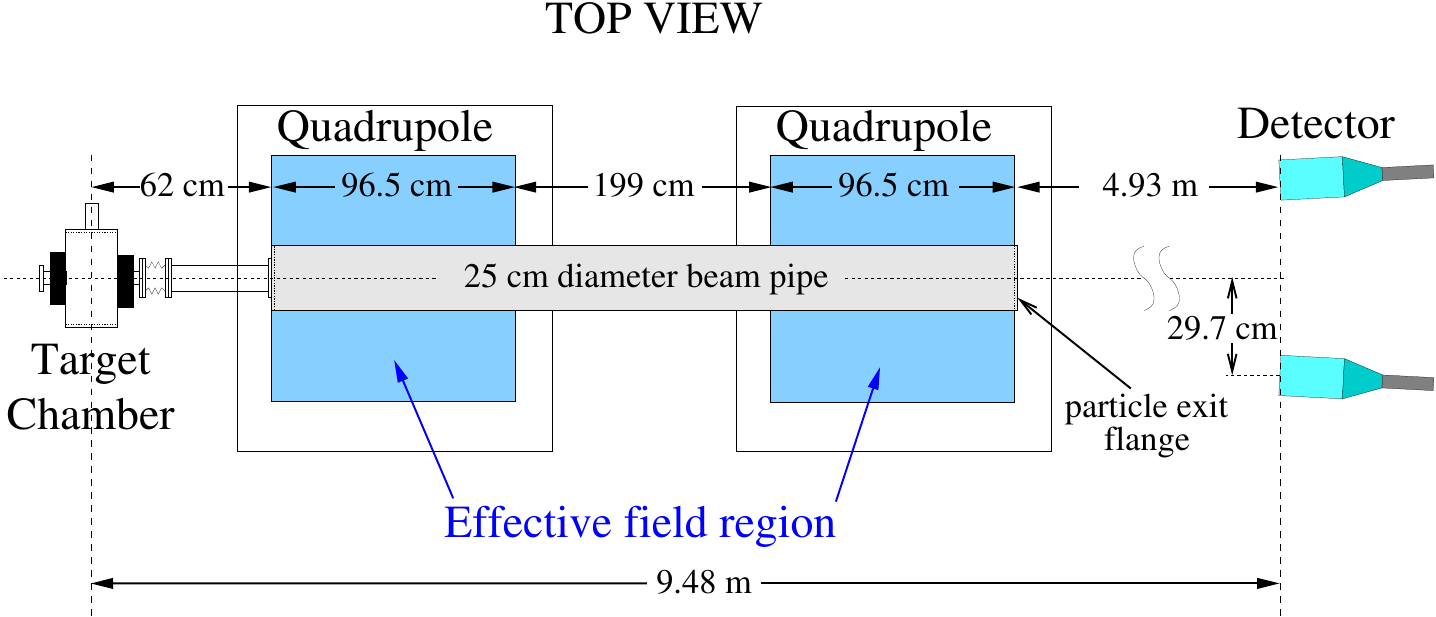}
\end{center}
\caption[]{Layout of the CLAS12 M{\o}ller polarimeter; detector shielding is not shown.}
\label{fig-PolLayout}
\end{figure}
The CLAS12 polarimeter, which schematic layout is shown on Fig.~\ref{fig-PolLayout}, relies upon a pair of quadrupole magnets to separate the scattered electrons from the beam and to deflect the scattered electrons into the detectors.
In Ref.~\cite{Gas17}, different options of positron beam polarization measurements in Hall-C are discussed. In particular their initial study show that no or relatively modest modification of the Hall-C M{\o}ller polarimeter will be needed  to allow 
measurements of the positron beam polarization. Hall-C and Hall-B M{\o}ller polarimeter layouts are very similar, and we expect same solutions will work in Hall-B as well.

\subsection{ Conicidence mode}
\begin{figure}[!htb]
\begin{center}
\includegraphics[width=0.425\textwidth]{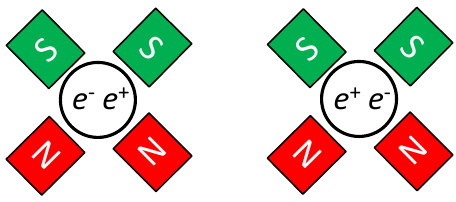}
\end{center}
\caption[]{M{\o}ller quadrupoles rewired into a dipole configuration as seen looking along the beam direction.  The left (right) panel has electrons (positrons) scattered to the left with positrons (electrons) scattered to the right.}
\label{fig-Dipoles}
\end{figure}

In the coincidence mode, both $e^{-}$ and $e^{+}$ are detected in coincidence, and hence they have very little background. With the positron beam, in order the scattered positron and the recoil electron to deflect into opposite directions as in the case of the M{\o}ller polarimeter, a dipole magnet(s) should be used instead of a quadrupole.
\begin{figure}[!htb]
    \centering
    \includegraphics[width=0.80\textwidth]{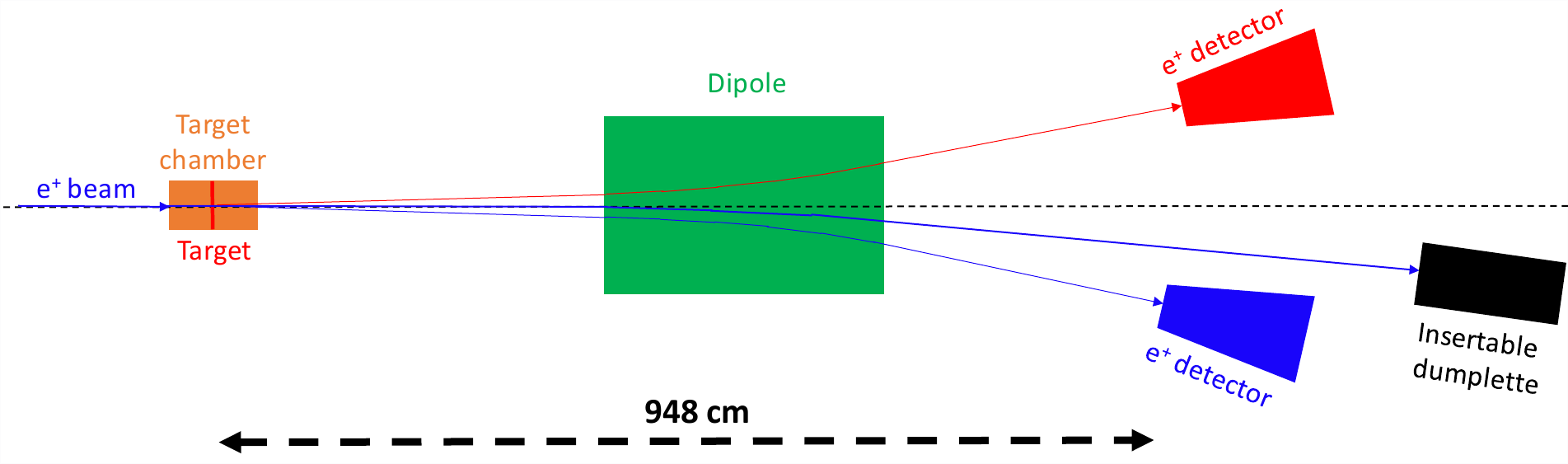} \\
    \includegraphics[width=0.80\textwidth]{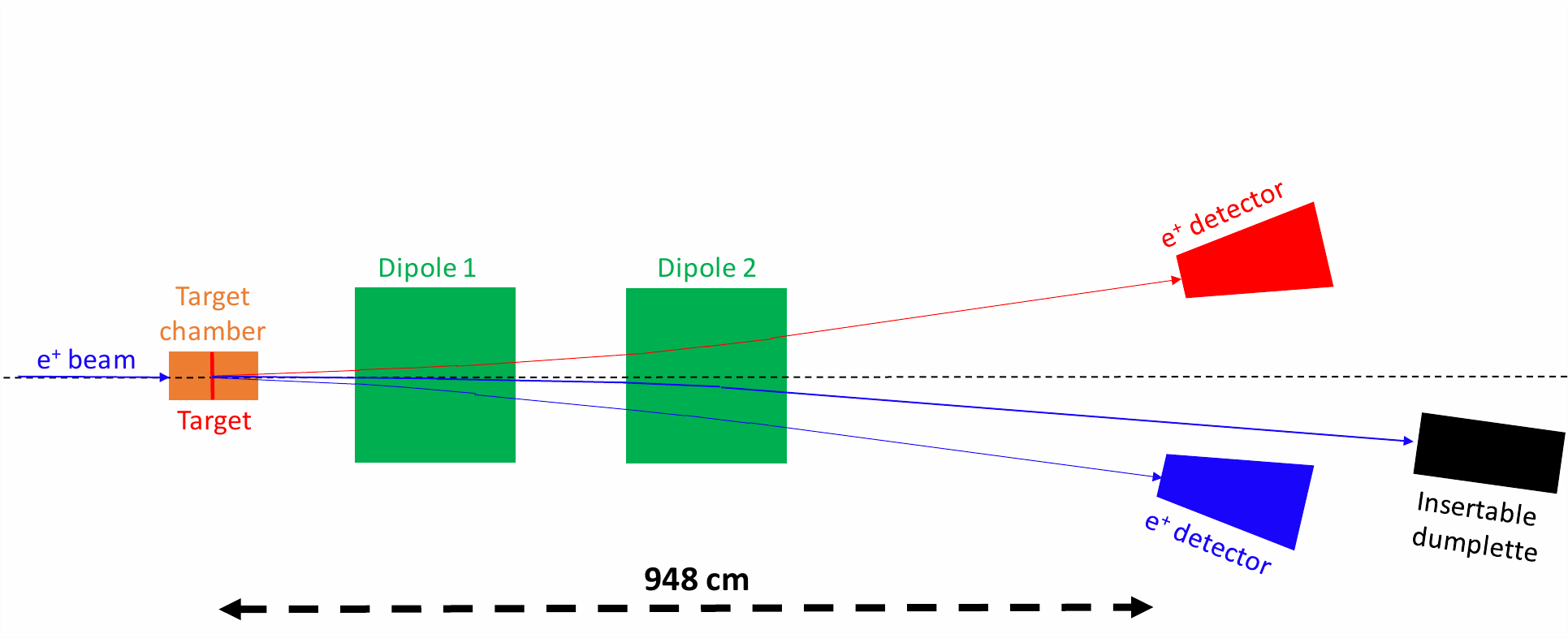}
    \caption{Schematic view of the Polarimeter. Top shows one dipole magnet configuration, while the bottom
    schematics shows the two dipole configuration.} 
    \label{fig:PolarimeterDipoleSchematic}
\end{figure}
In terms of a \emep pair detection, the dipole magnet is a good option, however the dipole magnet will deflect
the primary beam as well. The deflection can be significant enough for the beam to miss the tagger dump.
A workaround can be the installation of an insertable dumplette to be used during polarization measurements.
In order to change to a dipole configuration, either a new dipole magnet can be installed in place of existing quadrupoles, or the existing quadrupoles can be re-wired to act as a dipole (see Fig.\ref{fig-Dipoles}).
The schematic view of the abovementioned setups are shown in Fig.~\ref{fig:PolarimeterDipoleSchematic}.
Top figure represents single dipole configuration, and the bottom figure represents the two dipole configuration.

\subsection{Single arm measurement mode}
Another approach is to do a single arm-measurement.
In this case quadrupoles will not be modified, the beam can be directed to the tagger dump or to the tagger yoke (for energies above 6 GeV)~\cite{Bal20}.
With a single arm measurement mode, we will have two options for running quadrupoles:
 
 \begin{itemize}
     \item Option 1: Deflect negative particles away from the beam and focus positive particles towards the beam.
     \item Option 2: Deflect positive particles away from the beam and focus negative particles towards the beam.
 \end{itemize}

While a realistic GEANT4 simulation is needed before choosing the best option to run, it is expected that the option 1 will produce significantly lower background. The reason for this is the fact that high cross-section physics background processes e.g. Mott scattering, wide angle bremmstrahlung, produce high energy low angle positrons, part of which overlaps (in terms of energy and scattering angle) with positrons from Bhabha scattering and hence can reach the detector with option 2 settings. With option 1, all these positrons will be focused towards the beamline. Instead, electrons from the BhaBha scattering will reach the detector. 
In Ref.~\cite{Hau99} it is mentioned that during M{\o}ller polarimeter studies in Hall-C the dominating background of the single arm measurement originated from the Mott process. As the Hall C M{\o}ller polarimeter is very close to the Hall B one, we can assume at this stage that most of the background of a single arm measurement coming from the Mott process will be absent with option 2. 

While all these different possibilities will be thoroughly investigated, the {\it single mode} or {\it coincidence mode} have the advantage of operating for both electrons and positrons with only changing the magnet current polarity, an important feature for minimizing systematics of polarized BCA measurements. Beyond simulations, the experimental investigation of the {\it single mode} operation with the CEBAF electron beam may readily provide some answer about the feasibility of this option.

\end{appendix}                                                                                
%
%
%
%
%
%
%

\newpage

%
%

%
%

%
%
\end{document}